\documentclass[aps,pra,twocolumn,nofootinbib,floatfix,superscriptaddress]{revtex4-2}
\usepackage{amsmath,mleftright,mathtools}
\usepackage{bm}
\usepackage{stix,microtype}
\usepackage{graphicx}
\usepackage{xspace}
\usepackage{units}
\usepackage[colorlinks,linkcolor=blue,citecolor=blue,urlcolor=blue]{hyperref}
\usepackage[dvipsnames]{xcolor}
\usepackage{cleveref}
\usepackage{array}   
\newcolumntype{L}{>{$}c<{$}} 
\newcolumntype{C}{>{$}c<{$}} 
\usepackage{dcolumn}
\usepackage{bbm}

\newcommand{\be}{\begin{equation}}
\newcommand{\ee}{\end{equation}}
\newcommand{\bea}{\begin{eqnarray}}
\newcommand{\eea}{\end{eqnarray}}

\def\a{\alpha}
\def\b{\beta}
\def\g{\gamma}
\def\G{\Gamma}
\def\d{\delta}

\def\e{\epsilon}

\def\h{\eta}
\def\th{\theta}

\def\k{\kappa}
\def\l{\lambda}
\def\L{\Lambda}
\def\m{\mu}
\def\n{\nu}
\def\c{\xi}

\def\p{\pi}

\def\s{\sigma}
\def\S{\Sigma}
\def\t{\tau}
\def\f{\phi}
\def\vf{\varphi}
\def\F{\Phi}

\def\w{\omega}
\def\W{\Omega}

\def\bld{{\mathbf d}}
\def\ble{{\mathbf e}}

\def\blk{{\mathbf k}}

\def\blp{{\mathbf p}}
\def\blq{{\mathbf q}}
\def\blr{{\mathbf r}}

\def\blE{{\mathbf E}}

\def\blJ{{\mathbf J}}

\def\callA{\mbox{$\mathcal{A}$}}
\def\callB{\mbox{$\mathcal{B}$}}
\def\callC{\mbox{$\mathcal{C}$}}
\def\callD{\mbox{$\mathcal{D}$}}

\def\callT{\mbox{$\mathcal{T}$}}

\def\bcallC{\mbox{\boldmath $\mathcal{C}$}}

\def\ra{\rightarrow}

\def\iif{\infty}
\def\bra{\langle}
\def\ket{\rangle}
\def\grad{\mbox{\boldmath $\nabla$}}

\def\Tr{{\rm Tr}}
\def\Re{{\rm Re}}

\def\iu{{\rm i}}
\def\1op{\hat{\mathbbm{1}}}
\def\nn{\nonumber}

\begin{document}

\title{Non-Hermitian Bethe-Salpeter Equation for Open Systems: 
Emergence of Exceptional Points in Excitonic Spectra from First 
Principles}  

\author{Zhenlin Zhang}
\author{Wei Hu}
 \affiliation{
 Hefei National Research Center for Physical Sciences at the Microscale,\\ University of Science and Technology of China, Hefei, Anhui 230026, China}%

\author{Enrico Perfetto}%
\author{Gianluca Stefanucci}%
 \affiliation{%
 Dipartimento di Fisica, Università di Roma Tor Vergata, Via della 
 Ricerca Scientifica 1, 00133 Rome, Italy
}%
\affiliation{INFN, Sezione di Roma Tor Vergata, Via della Ricerca Scientifica 1, 00133 Rome, Italy}

\begin{abstract}
In open quantum systems hosting excitons, dissipation mechanisms 
critically shape the excitonic dynamics, band-structure and topological 
properties.  	
A microscopic understanding of excitons in such non-Hermitian 
settings demands a  first-principles 
generalization of the Bethe–Salpeter equation (BSE). 
Building on a recently 
introduced nonequilibrium Green’s function  formalism 
compatible with Lindbladian dynamics, we derive a non-Hermitian BSE 
from diagrammatic perturbation theory on the Keldysh contour, 
and obtain a microscopic  excitonic 
Hamiltonian that incorporates dissipation while preserving causality.            
We apply the formalism to valley excitons in transition metal 
dichalcogenides  coupled to structured photon baths. 
We uncover a rich landscape of exceptional points in momentum space, 
forming either discrete sets or continuous manifolds, depending 
on bath structure.   
The exceptional points give rise to nonanalytic valley-polarization, 
unusual polarization 
pattern in photoluminescence, and nontrivial topological 
signatures.   
Our results establish a first-principles framework for predicting and 
controlling excitonic behavior in open quantum materials, 
showing how engineered environments can be leveraged to induce 
and manipulate non-Hermitian and topological properties.

\end{abstract}

\maketitle

\section{Introduction}

Excitons, bound states of electrons and holes, play a central role in 
the optical and transport properties of semiconductors and 
insulators. In recent years, there has been growing interest in 
studying excitons in \textit{open systems}~\cite{mandal2020nonreciprocal,hu2023wave,hu2024generalized,Toffoletti_2025}, motivated by 
emerging experimental platforms such as exciton-polariton condensates 
in microcavities, excitonic devices coupled to lossy environments, 
and topological photonic 
structures~\cite{gan_controlling_2013,liu_strong_2015,gao2015observation,wurdack2023negative}. 
These systems, characterized by non-Hermitian dynamics, energy loss, 
and coupling to baths, open the door to novel quantum phenomena that 
do not exist in closed equilibrium 
settings~\cite{low_polaritons_2017,el-ganainy_non-hermitian_2018,ashida_non-hermitian_2020,bergholtz_exceptional_2021,roccati_non-hermitian_2022,ding_non-hermitian_2022,reitz_cooperative_2022,sieberer_universality_2023,fazio_many-body_2025,stefanini2025lindbladme}. Understanding how 
dissipation affects exciton dynamics, lifetimes, and even topological 
properties is both fundamentally intriguing and technologically 
relevant.

In the context of open systems, excitons have often been modeled as effective bosonic 
quasi-particles -- a framework that has proven powerful in capturing 
collective phenomena such as 
condensation~\cite{sieberer2013dynamical,smirnov2014dynamics}.  
These models, typically based on phenomenological approaches or 
effective Hamiltonians, provide valuable intuition and have been 
useful in advancing our understanding of many-body exciton 
physics. However, they generally abstract away the underlying 
fermionic nature of excitons and the microscopic origins of 
dissipation.   
As a result, while highly insightful, the effective bosonic 
description lacks the accuracy required for material-specific  
quantitative analysis.

A rigorous, first principles description of excitons is provided by 
the Bethe-Salpeter equation (BSE)~\cite{salpeter_relativistic_1951}, 
a Green’s function formalism built 
upon many-body perturbation 
theory~\cite{strinati_application_1988,RevModPhys.74.601,sander_beyond_2015,perfetto_nonequilibrium_2015,leng_gw_2016,blase_the-bethe-salpeter_2020}. The BSE 
has proven to be a powerful and accurate tool in describing excitonic 
effects in closed, equilibrium systems, particularly in the context 
of \textit{ab initio} calculations of solids and 
nanostructures.       

As early as 1966, Sham and Rice established a foundational link 
between exciton models and many-body theory~\cite{PhysRev.144.708}, 
recognizing the necessity of a field-theoretic treatment of excitonic 
phenomena. In 1974, Hanke and Sham performed the first 
non-model-based exciton calculations in solids using the 
BSE~\cite{PhysRevLett.33.582,PhysRevB.12.4501}, employing the 
time-dependent Hartree-Fock approximation and later incorporating 
static electron-hole screening~\cite{PhysRevB.21.4656}. A significant 
milestone was reached in 1995, when the Na$_4$ cluster became the 
first realistic system treated using a fully \textit{ab initio} BSE 
scheme~\cite{PhysRevLett.75.818}. Since then, the application of BSE 
has flourished, with widespread use in the quantitative prediction of 
optical properties of both finite and extended 
systems~\cite{schmidt_efficient_2003,MARINI20091392,DESLIPPE20121269,krause_implementation_2017,vorwerk_bethe-salpeter_2019,romero_abinit_2020,wallerberger_solving_2021}.                 

Despite these successes, the generalization of the BSE to open, 
dissipative systems remains lacking. In modern experimental 
platforms -- ranging from cavity-coupled materials to 
detuned or near-resonant driven systems -- loss, decoherence, and environmental 
coupling are intrinsic and unavoidable. These features fundamentally 
alter the nature of excitations, 
even in the weak light–matter coupling regime (i.e., without 
exciton–polariton formation)~\cite{wang2024non}. 
They introduce complex eigenvalues, 
non-Hermitian dynamics, and nonequilibrium steady states — all of 
which lie beyond the scope of traditional BSE treatments.

A suitable formalism to describe the aforementioned physics is 
the combination of Nonequilibrium Green's functions (NEGF) 
theory~\cite{svl-book_2025,kamenev_field_2011,stefanucci_in-and-out_2023} 
and Lindbladian 
dynamics~\cite{lindblad1976generators,Breuer_the_theory_2007}.
This has been achieved using field theoretical 
techniques~\cite{sieberer_keldysh_2016,fogedby_field-theoretical_2022,thomson_field_2023,sieberer_universality_2023}  
and the second-quantization 
approach~\cite{stefanucci2024kadanoff}. 
In this work, we build upon the formalism of 
Ref.~\cite{stefanucci2024kadanoff}, which is closer to the ab initio 
community and more suitable to study transient phenomena and 
relaxation dynamics of carriers, excitons and 
phonons~\cite{aoki_nonequilibrium_2014,MurakamiPRL2017,MolinaSanchez2017,tuovinen_comparing_2020,perfetto_real_2022,perfetto_real-time_2023}. By 
generalizing the BSE within this framework, we derive a first 
principles approach to treat excitons in dissipative environments and 
extract the microscopic non-Hermitian excitonic Hamiltonian. This 
development provides a powerful tool for quantitative predictions of 
excitonic phenomena beyond the limits of closed and Hermitian 
systems.

We also apply the formalism to study the non-Hermitian excitonic spectrum of 
two-dimensional semiconductors, such as transition metal 
dichalcogenides (TMDs). These systems feature  strong excitonic effects and 
high environmental sensitivity. Consequently, the optical properties 
of TMD -- whether suspended or or supported on substrates
-- can undergo significant modifications when 
embedded in a bath. Here we investigate two  
engineered baths: a bath of linearly polarized photons and a bath 
of circularly polarized photons carrying orbital angular 
momentum.  Depending on the bath, we find either a finite set or 
a continuous ring of exceptional points (EPs) in momentum space. 
These EPs give rise to nontrivial valley polarizations, 
photoluminescence features and topological properties.    
Our microscopic, non-Hermitian treatment highlights the emergence of 
rich excitonic physics even for weak exciton-photon coupling, and 
outlines a route toward controlling and tailoring excitonic behavior 
through bath engineering.       

The paper is organized as follows. In Section~\ref{ossec}, we briefly 
introduce the Lindblad equation that governs the dynamics of open 
quantum systems and outline the dissipation mechanisms considered in 
this work—namely, one-body loss and gain, as well as particle-hole 
recombination. In Section~\ref{Dissipative NEGF formalism}, we develop the 
dissipative nonequilibrium Green’s function (NEGF) formalism that 
arises from the underlying Lindbladian dynamics.     
In particular, we show that the problem can be reformulated as
a generalized diagrammatic perturbation theory on the Keldysh 
contour, made possible by a dissipative version of Wick’s theorem.
Foundational results for the one-particle NEGF are presented in 
Section~\ref{oneGsec}. We introduce 
the self-energy, address subtleties related to mean-field 
approximations, and present the framework for conserving 
approximations. These results are preparatory for deriving the BSE in 
an open setting. In Section~\ref{Dissipative BSE}, we establish the 
exact form of the non-Hermitian BSE for the 
four-time two-particle NEGF. By 
gradually introducing physically motivated approximations for the 
interaction kernel and the quasiparticle Green’s functions, we then derive 
a stationary BSE for the two-time response function.        
In Section~\ref{Semiconductors}, we specialize the stationary BSE 
formalism to semiconductors and derive the  non-Hermitian excitonic 
Hamiltonian. The microscopic derivation highlights that a Lindbladian 
treatment of dissipation imposes constraints on the allowable 
dissipation channels: arbitrary choices can either violate the 
stationarity assumption or compromise causality.    
Physically relevant and formally admissible dissipation mechanisms 
are explicitly discussed.
In Section~\ref{Valley excitons in TMD}
we apply the formalism to valley excitons in  
TMDs embedded in different photon baths. 
We present a $2\times 2$ effective non-Hermitian description of the excitonic 
subspace and analyze its eigenvalues and eigenvectors, revealing the 
emergence of EPs. The occurrence of EPs induces a nonanalytic behavior 
in the valley polarization, and leads to a complex polarization 
pattern in the photoluminescence emission.    
Finally, we compute the Berry curvature and report a nontrivial 
topological character of the non-Hermitian excitons.
A summary of the main results and an outlook for future directions 
are presented in Section~\ref{concsec}.                               

\section{The open system}
\label{ossec}

The time evolution for the reduced many-body 
density matrix $\hat{\rho}(t)$ of an open quantum system governed 
by a Lindbladian dynamics reads~\cite{10.1063/1.522979,lindblad1976generators}:
\begin{align}\label{lindblad eq}
    \frac{\rm d}{{\rm d} t}\hat{\rho}(t)=
	-{\rm i}\left[\hat{H}(t),\hat{\rho}(t)\right]_-+
	\hat{\callD}[\hat{\rho}](t).
\end{align}
The coherent (unitary) evolution is 
determined by the many-body Hamiltonian of the closed system
\begin{align}
    \hat{H}(t)=\sum_{ij}h_{ij}(t)\hat{d}_i^\dagger\hat{d}_j+
	\frac{1}{2}\sum_{ijmn}v^{\rm phys}_{ijmn}(t)
	\hat{d}_i^\dagger\hat{d}_j^\dagger\hat{d}_m\hat{d}_n,
\label{MBham}
\end{align}
where the annihilation operators $\hat{d}_i$ are either bosonic or 
fermionic, obeying the commutation or anticommutation rules 
$[\hat{d}_i,\hat{d}^\dagger_j]_\mp =\d_{ij}$ (upper sign for bosons 
and lower sign for fermions). The global $U(1)$ symmetry, $\hat{d}_j \ra \hat{d}_j 
e^{{\rm i}\a}$, of $\hat{H}(t)$ guarantees that the total number of 
particles of the closed system is conserved.       

The one-particle Hamiltonian $h(t)$ comprises 
the kinetic energy as well as static or time-dependent external fields.
The time-dependence of the physical interaction $v^{\rm phys}$ 
could describe, e.g., 
the adiabatic switching of the Coulomb interaction for electronic system or 
the quenching of the dipolar interaction in ultra-cold atomic and molecular 
systems~\cite{6b82-sb2j}.

The dissipation induced by the environment is described by 
a set of so-called dissipator superoperators
\begin{align}
    \hat{\callD}[\hat{\rho}](t)=\sum_\l \left[\hat{\callD}^{\rm p}_{\l}[\hat{\rho}](t)+\hat{\callD}^{\rm 
	h}_{\l}[\hat{\rho}](t)+\hat{\callD}^{\rm ph}_{\l}[\hat{\rho}](t)\right],
\end{align}
which represent the different kind of dissipation  considered in this 
work -- one-body loss (p), one-body gain (h), and particle-hole 
recombination (ph)
-- in multi channels indexed by $\l$. For a many-body system, there are typically several decay channels, e.g., when the system is coupled to several baths. 

The dissipators consist of a jump term and an anticommutator
\begin{align}\label{dissipatorD}
    \hat{\callD}^{\k}_\l[\hat{\rho}](t)=
	2\hat{L}^{\k}_{\l}(t)\hat{\rho}(t)\hat{L}_{\l}^{\k\dagger}(t)
	-\left[\hat{L}^{\k\dagger}_{\l}(t)\hat{L}^{\k}_{\l}(t),\hat{\rho}(t)\right]_+,
\end{align}
where the quantum jump operators   
$\hat{L}^{\k}_{\l}$ of type ${\k}={\rm p},{\rm h},{\rm ph}$
describe the interaction between the system and the environment.

It is worth recalling that deriving the Lindblad equation from the 
underlying system-environment Hamiltonian generally leads to 
non-Hermitian jump operators.  If the jump operators are 
Hermitian and conserved by the coherent dynamics 
($[\hat{L}^{\k}_\l,\hat{H}]_-=0$), then they can only induce pure 
decoherence~\cite{sieberer_universality_2023,stefanini2025lindbladme}. 
As for Hermitian jump operators that do not commute with the 
physical Hamiltonian, they drive the system toward an 
infinite-temperature  
state~\cite{stefanini2025lindbladme,picano2025heatingdynamicscorrelatedfermions,penc2025linearresponseexacthydrodynamic,Anonymous_2025}.

The most general form of the jump operators in 
Eq.~(\ref{dissipatorD}) is
\begin{subequations}
\begin{align}
    \hat{L}_{\l}^{\rm p}(t)&=\sum_n \eta^{\l}_n(t)\hat{d}_n,\label{loss}\\ 
    \hat{L}_{\l}^{\rm h}(t)&=\sum_n\bar{\h}^{\l *}_n(t)\hat{d}^\dagger_n,\label{gain}\\ 
    \hat{L}_{\l}^{\rm ph}(t)&=\sum_{mn} \c_{mn}^{\l}(t) \hat{d}_m^\dagger \hat{d}_n,\label{phloss}
\end{align}
\label{genlind}
\end{subequations}
where the coefficients $\h$, $\bar{\h}$ and $\c$ determine the 
dissipation rates due to one-body loss, one-body gain and 
particle-hole recombination. 

The symmetries preserved in Lindbladian dynamics can be classified 
into two categories: \textit{weak} and \textit{strong}. A symmetry is 
said to be weak if both the Hamiltonian $\hat H$ and
the dissipators $\hat{\mathcal{D}}$ 
remain invariant under the corresponding symmetry transformations, 
even though the jump operators $\hat{L}$ themselves do not. In 
contrast, a weak symmetry is considered strong when the 
jump operators themselves are invariant under the 
symmetry transformations.         

The forms of the one-body jump operators ($\k={\rm p,h}$)
imply that the corresponding 
dissipators have $weak$ global $U(1)$ symmetry: the total particle 
number is $not$ conserved, yet the many-body density matrix 
$\hat{\rho}$ does not develop quantum coherence between 
Hilbert space sectors of different particle 
number~\cite{sieberer_universality_2023,deGroot2022symmetryprotected,PhysRevLett.125.240405,mittal2025fermionquantumcriticalityfar,chaduteau2025lindbladianversuspostselectednonhermitian,PhysRevLett.128.033602,Buca_2012,PhysRevA.89.022118}.             
The one-body jump operators are under intense investigation in the 
community of, e.g., non-Hermitian skin 
effect~\cite{yokomizo2019non,chaduteau2025lindbladianversuspostselectednonhermitian} and topological superconductor~\cite{gogoi2025dissipationinducedmajarona0,PhysRevB.101.014306,ghosh2022non}.      

The particle-hole jump operator, quadratic in field operators, 
possess a $strong$ global $U(1)$ symmetry -- both $\hat{\callD}^{\rm 
ph}$ and $\hat{L}^{\rm ph}$ are invariant under global $U(1)$ 
transformation -- and renormalizes both the 
single particle Hamiltonian and the physical interaction 
during time evolution. Thus, analytical solutions are in general 
unavailable even for $v^{\rm phys}=0$.
Particle-hole losses have drawn special attention in the context of 
critical phenomena   out of 
equilibrium~\cite{mittal2025fermionquantumcriticalityfar} and novel 
quantum platform such as 
exciton-polariton~\cite{carusotto2013quantum,PhysRevLett.99.140402} 
and disordered 
systems~\cite{feng2025homogeneoussystemsmeetdissipation,PhysRevLett.132.216301}.

\section{Dissipative NEGF formalism}
\label{Dissipative NEGF formalism}

\subsection{Hamiltonian and  $n$-particle Green's functions on the Keldysh contour}  

The $n$-particle nonequilibrium Green's function (NEGF) of the open system 
governed by the Lindbladian dynamics can be written as~\cite{stefanucci2024kadanoff} 
\begin{align}\label{n-particleNEGF}
    &G_n(j_1z_1,\ldots,j_nz_n;j_1'z_1',\ldots,j_n'z_n')\nn\\
    &\equiv\frac{1}{{\rm i}^n}
    \big\bra\hat{d}_{j_1}(z_1)
	\ldots\hat{d}_{j_n}(z_n)\hat{d}^\dagger_{j_n'}(z_n'^+)
	\ldots\hat{d}^\dagger_{j_1'}(z_1'^+)\big\ket,
\end{align}
where the times of the field operators run
along the oriented Keldysh contour $C=C_- \cup C_+=(t_0,\infty)\cup 
(\infty,t_0)$, and we use the notation $z^{+}$ to denote a time  
infinitesimally later than $z$ on $C$.
In Eq.~(\ref{n-particleNEGF}) we use
the shorthand notation    
\begin{align}\label{shorthand}
    \big\bra\ldots\big\ket\equiv \Tr \big[\hat{\rho}(t_0)\callT\big\{
    e^{-{\rm i}\int_C {\rm d}\bar{z}\hat{\mathcal{H}}(\bar{z})}\ldots
    \big\}\big],
\end{align}
where $\callT$ is the contour ordering operator, and
the dissipative Hamiltonian generator $\hat{\mathcal{H}}(z)$ 
reads
\begin{align}
\hat{\mathcal{H}}(z)&=\hat{H}(z)+\sum_{\k,\l}\Big[
-{\rm i}s(z)
\hat{L}^{\k\dagger}_{\l}(z)\hat{L}^{\k}_{\l}(z)
\nn\\
&+{\rm i}\th_{-}(z)
\hat{L}^{\k\dagger}_{\l}(z^{\ast})\hat{L}^{\k}_{\l}(z)
-{\rm i}\th_{+}(z)
\hat{L}^{\k\dagger}_{\l}(z)\hat{L}^{\k}_{\l}(z^{\ast})
\Big].
\label{hzatez}
\end{align}

Let us discuss in some detail 
the Hamiltonian in Eq.~(\ref{hzatez}).
We write a  time $z\in C$ as  $z=t_\pm$ if $z\in 
C_\pm$. The operators 
$\hat{H}(z=t_{\pm})=\hat{H}(t)$ and 
$\hat{L}^{\k}_{\l}(z=t_{\pm})=\hat{L}^{\k}_{\l}(t)$ are the 
same on both branches of the contour. The functions 
$\th_\pm(z)=1$ if $z\in C_\pm$ and zero otherwise, whereas 
$s(z)=\th_-(z)-\th_+(z)$ is a sign function. Finally, we 
define $z^*=t_\mp$ if $z=t_\pm$. Although 
the dissipative Hamiltonian $\hat{\mathcal{H}}(z)$ depends on both 
$z$ and $z^{\ast}$, it is easy to verify that 
$\hat{\mathcal{H}}(t_{-})=\hat{\mathcal{H}}(t_{+})=\hat{H}(t)$, and 
therefore the operator $\hat{K}(t)\equiv 
\callT\big\{e^{-{\rm i}\int_{t_-}^{t_+}{\rm 
d}\bar{z}\hat{\mathcal{H}}(\bar{z})}\big\}=1$ for all 
$t$, reflecting the conservation of probability.

The 
dissipative Hamiltonian can be split into a noninteracting part 
$\hat{\mathcal{H}}_0(z)$ and interacting part 
$\hat{\mathcal{H}}_{\rm int}(z)$            
\begin{align}
    &\hat{\mathcal{H}}(\bar{z})
    =\hat{\mathcal{H}}_0(\bar{z})+\hat{\mathcal{H}}_{\rm int}(\bar{z}),
\label{splitham}
\end{align}
which are, respectively, quadratic in the field operators,
\begin{align}
    &\hat{\mathcal{H}}_0(\bar{z})
    =\sum_{ij}\int_C {\rm d}\bar{z}' 
	h_{ij}(\bar{z}',\bar{z})\hat{d}_i^\dagger 
	(\bar{z}')\hat{d}_j(\bar{z}),
\label{onebodyH}
\end{align}
and quartic in the  field operators
\begin{align}
    \hat{\mathcal{H}}_{\rm int}(\bar{z})
    =\frac{1}{2}\sum_{ijmn}&\int_C {\rm d}\bar{z}' v_{ijmn}(\bar{z},\bar{z}')\nn\\
    &\times\hat{d}_i^\dagger(\bar{z}^{\prime +})\hat{d}_j^\dagger(\bar{z}^+)
	\hat{d}_m(\bar{z})\hat{d}_n(\bar{z}').
\label{twobodyH}
\end{align}
The explicit forms of Eqs.~(\ref{onebodyH}) and~(\ref{twobodyH}) were 
derived in Ref.~\cite{stefanucci2024kadanoff}. For completeness, and to 
ensure the manuscript is self-contained, a more detailed derivation 
is provided in Appendix~\ref{lindHapp}. Below, we summarize the main 
results.     

The noninteracting one-particle 
Hamiltonian reads (upper/lower sign for bosons/fermions)        
\begin{align}
    h_{ij}(z',z)&=\d(z',z)h_{ij}(t)\nn\\
    &-{\rm i}s(z)\d(z',z)\Big(V_{ij}^{\rm loss}(t)\pm 
    V_{ij}^{\rm gain}(t)+ V_{ij}^{\rm diss}(t)\Big)\nn\\
    &+2{\rm i}\d(z',z^*)\Big(\th_-(z)V_{ij}^{\rm loss}(t)\mp\th_+(z) 
	V_{ij}^{\rm gain}(t)\Big),
\label{oneph}
\end{align}
where we introduce the Dirac delta function on the contour, 
i.e., $\int_C {\rm d}z'\d(z,z')f(z')=f(z)$ for any function $f$. 
In addition to the physical one-particle Hamiltonian $h_{ij}$ of 
Eq.~(\ref{MBham}), the noninteracting one-particle 
Hamiltonian of Eq.~(\ref{oneph}) contains complex potentials due to  one-body 
loss and gain mechanisms, i.e.,
\begin{subequations}
\begin{align}
    V_{ij}^{\rm loss}(t)&=\sum_{\l} \h_i^{\l *}(t)\h_j^{\l}(t),
	\label{Vloss}\\
    V_{ij}^{\rm gain}(t)&=\sum_{\l} \bar{\h}_j^{\l *}(t)\bar{\h}_i^{\l}(t),
	\label{Vgain}
\end{align}
\end{subequations}
as well as from particle-hole recombination mechanisms, i.e., 
\begin{align}
V_{ij}^{\rm diss}(t)= \frac{1}{2}\sum_m v_{imjm}^{\rm diss}(t)
\label{cpvdiss1}
\end{align}
with
\begin{align}
v_{ijmn}^{\rm diss}(t)&=2\sum_{\l}\c^{\l *}_{ni}(t)\c^{\l}_{jm}(t).
\label{dissint}
\end{align}	

The generalized interaction in Eq.~(\ref{twobodyH}) is the sum of the physical 
interaction and the dissipation-induced interaction driven by the 
particle-hole jump operators
\begin{align}
    v_{ijmn}(z,z')&=v_{ijmn}^{\rm phys}(z,z')+v_{ijmn}^{\rm diss}(z,z')\label{generalized v}\\
    v_{ijmn}^{\rm phys}(z,z')&=\d(z,z')v_{ijmn}^{\rm phys}(t)\label{vphys}\\
    v_{ijmn}^{\rm diss}(z,z')&=-{\rm i}s(z)\left[\d(z,z')+\d(z,z^{\prime 
	*})\right]v_{ijmn}^{\rm s,diss}(t)\nn\\
    &-{\rm i}\d(z,z'^*)v_{ijmn}^{\rm a,diss}(t)\label{vdiss},
\end{align}
where the symmetric ($v^{\rm s,diss}$) and antisymmetric ($v^{\rm a,diss}$) tensors
are defined in terms of the tensor introduced in Eq.~(\ref{dissint}):
\begin{subequations}
\begin{align}
    v_{ijmn}^{\rm s,diss}(t)&=\frac{1}{2}
	\left[v_{ijmn}^{\rm diss}(t)+ v_{jinm}^{\rm diss}(t)\right],
	\\
	    v_{ijmn}^{\rm a,diss}(t)&=\frac{1}{2}
	\left[v_{ijmn}^{\rm diss}(t)- v_{jinm}^{\rm diss}(t)\right].
\end{align}
\label{vphsa}
\end{subequations}

The generalized interaction 
preserves the symmetry property 
\begin{align}
	v_{ijmn}(z,z')=v_{jinm}(z',z)
	\label{symkv}
\end{align}
of the physical interaction, 
an essential feature for deriving the prefactors of the Feynman 
diagrams~\cite{stefanucci2024kadanoff}. If the particle-hole jump 
operators are chosen to be either Hermitian, i.e., 
$\c_{nm}^{\l *}(t)= \c_{mn}^{\l }(t)$, 
or anti-Hermitian, i.e., $\c_{nm}^{\l *}(t)=- \c_{mn}^{\l 
}(t)$, then the antisymmetric part vanishes ($v^{\rm a,diss}=0$).
The many-body Hamiltonian of the closed system is recovered by 
setting  all the Lindblad coefficients to zero, as it should be.

Henceforth, to avoid ambiguity in expressions involving 
$\pm$, the upper (lower) sign refers to bosons (fermions). When 
the sign appears as a subscript on time variables, e.g.,   $t_\pm\in 
C_\pm$, it denotes the corresponding branch of the Keldysh contour.           

\subsection{Diagrammatic expansions}

Differentiating the $n$-particle NEGF in Eq.~(\ref{n-particleNEGF}) with 
respect to $z$ or $z'$ gives rise to the so called Martin-Schwinger 
hierarchy (MSH)~\cite{PhysRev.115.1342,svl-book_2025}, which couples $G_{n}$
to the $(n+1)$ and 
$(n-1)$-particle NEGFs.  In Appendix~\ref{MSapp} we 
prove that the MSH for our choice of jump operators  reads
\begin{widetext}
\begin{align}
    &\sum_{\bar{j}}\int_C {\rm d}\bar{z}\left[{\rm i}\frac{{\rm d}}{{\rm d}z_q}\d_{j_q\bar{j}}\d(z_q,\bar{z})-h_{j_q\bar{j}}(z_q,\bar{z})\right]G_n(j_1z_1,\ldots ,\bar{j}\bar{z},\ldots ,j_nz_n;j_1'z_1',\ldots,j_n'z_n')\nn\\
    &=\sum_{l=1}^n(\pm)^{q+l}\d_{j_qj_l'}\d(z_q,z_l')G_{n-1}(j_1z_1,\ldots ,\stackrel{\sqcap}{j_qz_q},\ldots ,j_nz_n;j_1'z_1',\ldots ,\stackrel{\sqcap}{j_l'z_l'},\ldots ,j_n'z_n')\nn\\
    &\pm {\rm i}\sum_{prs}\int_C {\rm d}\bar{z}v_{j_qprs}(\bar{z},z_q)G_{n+1}(j_1z_1,\ldots ,r\bar{z},sz_q,\ldots ,j_nz_n;j_1'z_1',\ldots,p\bar{z}^+,j_q'z_q',\ldots,j_n'z_n'),
\label{mshL}
\end{align}
and 
\begin{align}
    &\sum_{\bar{j}'}\int_C {\rm d}\bar{z}'G_n(j_1z_1,\ldots 
	,j_nz_n;j_1'z_1',\ldots 
	,\bar{j}'\bar{z}',\ldots,j_n'z_n')\left[\frac{1}{{\rm i}}\frac{\overleftarrow{{\rm d}}}{{\rm d}z_q'}\d_{\bar{j}'j_q'}\d(\bar{z}',z_q')-h_{\bar{j}'j_q'}(\bar{z}',z_q')\right]\nn\\
    &=\sum_{l=1}^n(\pm)^{q+l}\d_{j_lj_q'}\d(z_l,z_q')G_{n-1}(j_1z_1,\ldots ,\stackrel{\sqcap}{j_lz_l},\ldots ,j_nz_n;j_1'z_1',\ldots ,\stackrel{\sqcap}{j_q'z_q'},\ldots ,j_n'z_n')\nn\\
    &\pm {\rm i}\sum_{prs}\int_C {\rm d}\bar{z}' v_{srpj_q'}(\bar{z}',z_q')G_{n+1}(j_1z_1 ,\ldots,p\bar{z}'^+,j_qz_q,\ldots ,j_nz_n;j_1'z_1',\ldots ,r\bar{z}',sz_q',\ldots,j_n'z_n'),
\end{align}
\end{widetext}
where the symbol ${\sqcap}$ above an index signifies that the index is missing.
By setting the general interaction $v(z,z')$ to zero the 
$n$-particle NEGF couples exclusively to the 
$(n-1)$-particle NEGF, and the solution $G_n^0$ of the  
noninteracting MSH is equivalent to Wick's 
theorem~\cite{stefanucci2024kadanoff}. 

By construction, the noninteracting NEGF $G_n^0$ can alternatively be written as in 
Eq.~(\ref{n-particleNEGF}) with the replacement 
$\hat{\mathcal{H}}(z)\to \hat{\mathcal{H}}_0(z)$. Therefore, $G_n^0$ 
fully accounts for one-body loss and gain mechanisms. In the presence of 
particle-hole recombination, $\hat{\mathcal{H}}_0(z)$ also contains  the  
complex potential $V^{\rm diss}$ in Eq.~(\ref{cpvdiss1}). 
Due to this term the contour integral $\int_C {\rm d}z {\rm 
d}z' h(z,z')\neq 0$, and consequently the 
operator $\hat{K}_0(t)\equiv \callT\big\{e^{-{\rm i}\int_{t_-}^{t_+}{\rm 
d}\bar{z}\hat{\mathcal{H}}_0(\bar{z})}\big\}\neq 1$. 
This fact implies that all $G_n^0$ diverge in the long 
time limit. As we show below, the physical stability is restored to 
lowest order in perturbation theory, i.e., in the mean-field  Hartree-Fock 
(HF) approximation~\cite{stefanucci2024kadanoff}.
Rigorously speaking, the 
Wick's theorem can be proven only for the mean-field MSH.              

Expanding Eq.~(\ref{n-particleNEGF}) in powers of $\hat{\mathcal{H}}_{\rm 
int}(z)$ and using Wick's theorem, each term of the expansion can be 
represented by a Feynman diagram.
The terms 
containing vacuum diagrams cancel with the expansion of 
the denominator $1=\Tr[\hat{\rho}(t_0) \hat{K}(t_0)]$. Furthermore, the 
number of topological equivalent diagrams of order $k$ is the same as 
for physical interactions, i.e., $2^kk!$, since the generalized 
interaction has the same symmetry properties, see Eq.~(\ref{symkv}).
Hence, it is enough to consider only 
connected and topologically inequivalent diagrams. The final 
results for the one-particle and two-particle NEGFs are 
(integrals are over $z_1,z_1',\ldots,z_k,z_k'$ and repeated indices are summed 
over):                
\begin{align}\label{1-particle NEGF}
    &G_{ab}(z_a,z_b)\nn\\
    &=\sum_{k=0}^{\infty} {\rm i}^k\int_C v_{i_1j_1m_1n_1}(z_1,z_1')\ldots v_{i_kj_km_kn_k}(z_k,z_k')\nn\\
    &\times\left|\begin{array}{cccc}
G^0_{ab}(z_a , z_b) & G^0_{aj_1}(z_a , z_1^+) & \ldots & G^0_{ai_k}(z_a , z_k'^+) \\
G^0_{m_1b}(z_1 , z_b) & G^0_{m_1j_1}(z_1 , z_1^+) & \ldots & G^0_{m_1i_k}(z_1 , z_k'^+) \\
\vdots & \vdots & \ddots& \vdots \\
G^0_{n_kb}(z_k' , z_b) & G^0_{n_kj_1}(z_k' , z_1^+) & \ldots & G^0_{n_ki_k}(z_k' , z_k'^+) 
\end{array}\right|_{\substack{ \pm \\c\\{ t.i. }}},
\end{align}

\begin{align}\label{2-particle NEGF}
    &G_2(az_a,bz_b;cz_c,dz_d)\nn\\
    &=\sum_{k=0}^\infty {\rm i}^k\int_C v_{i_1j_1m_1n_1}(z_1,z_1')\ldots v_{i_kj_km_kn_k}(z_k,z_k')\nn\\
    &\times\left|\begin{array}{cccc}
G^0_{ac}(z_a , z_c) & G^0_{ad}(z_a , z_d) & \ldots & G^0_{ai_k}(z_a , z_k'^+) \\
G^0_{bc}(z_b , z_c) & G^0_{bd}(z_b , z_d) & \ldots & G^0_{bi_k}(z_b , z_k'^+) \\
\vdots & \vdots & \ddots& \vdots \\
G^0_{n_kc}(z_k' , z_c) & G^0_{n_kd}(z_k' , z_d) & \ldots & G^0_{n_ki_k}(z_k' , z_k'^+) 
\end{array}\right|_{\substack{ \pm \\c\\{ t.i. }}},
\end{align}
and the like for higher order NEGF. In the above equations 
\begin{align}
    G^{0}_{jj'}(z_j,z_j')\equiv G^{0}_1(jz_j;j'z_j')
\end{align}
is the noninteracting one-particle NEGF, and 
the labels ``\textit{c}'' and ``\textit{t.i.}'' indicate that in 
the permanent/determinat 
only connected and topologically inequivalent diagrams are 
retained.

\section{One-particle Green's function}
\label{oneGsec}
Before delving into the Bethe-Salpeter equation for the two-particle 
Green's function we need to introduce a few concepts emerging from 
the analysis of  the 
one-particle Green's function.

\subsection{Kadanoff-Baym equations}

The Wick's expansion of the one-particle NEGF provides the Feynman 
rules  for extracting the irreducible self-energy $\Sigma=\S[G^0,v]$ 
depending on the noninteracting $G^0$ and the generalized 
interaction $v$.
The full set of $\S$-diagrams can alternatively be obtained by 
dressing  $G^0$ and discarding the nonskeletonic diagrams 
(a diagram is skeletonic if it does not contain self-energy 
insertions)~\cite{svl-book_2025}.
Denoting by $\S^{\rm s}$ the sum of only skeletonic $\S$-diagrams, 
we have $\S=\S[G^0,v]=\S^{\rm s}[G,v]$. The equations of motion for 
$G$ can then be written as (in matrix form)
\begin{subequations}
\begin{align}
    {\rm i}\frac{{\rm d}}{{\rm d}z}&G(z,z')-\int_C {\rm d}\bar{z} 
	\left[h(z,\bar{z})+\Sigma^{\rm 
	s}(z,\bar{z})\right]G(\bar{z},z')=\d(z,z'),
	\label{eom1}\\
    \frac{1}{{\rm i}}\frac{{\rm d}}{{\rm d}z'}&G(z,z')-\int_C {\rm 
	d}\bar{z} G(z,\bar{z})\left[h(\bar{z},z')+\Sigma^{\rm 
	s}(\bar{z},z')\right]=\d(z,z').
	\label{eom2}
\end{align}
\label{eom}
\end{subequations}

Let us separate out the Hartree-Fock (HF) contribution 
from the full self-energy, i.e., $\S^{\rm s}=\S^{\rm 
HF}+\S^{\rm c}$, where $\S^{\rm HF}=\S^{\rm H}+\S^{\rm F}$ is the sum 
of the tadpole diagram    
\begin{align}\label{hartree}
    \S^{\rm H}_{in}(z,z')&=\pm {\rm i}\d(z,z')\sum_{mj}\int_C {\rm d}\bar{z}v_{ijmn}(\bar{z},z)G_{mj}(\bar{z},\bar{z}^+)\nn\\
    &= \d(z,z')\sum_{mj}\left(v_{ijmn}^{\rm phys}(t)-{\rm i}v_{ijmn}^{\rm a,diss}(t)\right)\rho_{mj}^<(t),
\end{align}
where $\rho^{<}(t)\equiv \pm {\rm i}G(z,z^{+})$ is the 
one-particle density matrix,
and oyster diagrams
\begin{align}\label{fock}
    \S^{\rm F}_{in}(z,z')={\rm i}\sum_{mj}G_{mj}(z,z')v_{ijnm}(z',z).
\end{align}
The time local structure of the generalized interaction in 
Eqs.~(\ref{generalized v}-\ref{vdiss}) implies that the HF 
self-energy  effectively renormalizes the noninteracting one-particle 
Hamiltonian $h(z,z')$ in Eq.~(\ref{oneph}), and stabilizes  
the system dynamics since $\int_C {\rm d}z {\rm d}z'[h(z,z')+\S^{\rm 
HF}(z,z')]=0$, see Appendix~\ref{HFapp}. Notice that the quantity
\begin{align}
V^{\rm a,diss}_{in}(t)\equiv -{\rm i}\sum_{mj}v_{ijmn}^{\rm 
a,diss}(t)\rho_{mj}^<(t)
\label{R}
\end{align}
in the Hartree self-energy is a Hermitian matrix and therefore it 
provides a dissipation-induced renormalization of the one-particle 
energy levels.

The dissipative Kadanoff-Baym equations (KBE) are equations of motion 
for the Keldysh components of the NEGF $G(z,z')$ with times on the 
contour. Setting $z=t_{\mp}$ and $z'=t'_\pm$ we have the lesser and 
greater component 
$G^{\lessgtr}(t,t')\equiv G(t_{\mp},t'_\pm)$. Using the Langreth 
rules~\cite{langreth1976linear,svl-book_2025} to evaluate the 
convolutions in Eqs.~(\ref{eom}) we find~\cite{stefanucci2024kadanoff,blommel_unified_2025}    
\begin{subequations}
\begin{align}
    &\left[{\rm i}\frac{{\rm d}}{{\rm d}t}-h^{\rm HF}(t)\right]G^<(t,t')\pm 2{\rm i}{\g}^<(t)G^{\rm A}(t,t')\nn\\
    &=\int_{t_0}^\infty {\rm d}\bar{t}\left[\Sigma^{\rm c,<}(t,\bar{t}) G^{\rm A}(\bar{t},t')+\Sigma^{\rm c,R} (t,\bar{t})G^{<} (\bar{t},t')\right],\label{kbe1}\\
    &\left[{\rm i}\frac{{\rm d}}{{\rm d}t}-h^{\rm HF}(t)\right]G^>(t,t')+ 2{\rm i}{\g}^>(t)G^{\rm A}(t,t')\nn\\
    &=\int_{t_0}^\infty {\rm d}\bar{t}\left[\Sigma^{\rm c,>}(t,\bar{t}) G^{\rm A}(\bar{t},t')+\Sigma^{\rm c,R} (t,\bar{t})G^{>} (\bar{t},t')\right],\label{kbe2}\\
    &G^<(t,t')\left[\frac{1}{{\rm i}}\frac{\overleftarrow{\rm d}}{{\rm d}t'}-h^{\rm HF\dagger}(t')\right]\pm 2{\rm i}G^{\rm R}(t,t'){\g}^<(t')\nn\\
    &=\int_{t_0}^\infty {\rm d}\bar{t}\left[G^{<}(t,\bar{t}) \Sigma^{\rm c,A}(\bar{t},t')+G^{\rm R} (t,\bar{t}) \Sigma^{\rm c,<}(\bar{t},t') \right],\label{kbe3}\\
    &G^>(t,t')\left[\frac{1}{{\rm i}}\frac{\overleftarrow{\rm d}}{{\rm d}t'}-h^{\rm HF\dagger}(t')\right]+ 2{\rm i}G^{\rm R}(t,t'){\g}^>(t')\nn\\
    &=\int_{t_0}^\infty {\rm d}\bar{t}\left[G^{>}(t,\bar{t}) \Sigma^{\rm c,A}(\bar{t},t')+G^{\rm R} (t,\bar{t}) \Sigma^{\rm c,>}(\bar{t},t') \right],\label{kbe4}
\end{align}
\label{kbe}
\end{subequations}
where the retarded and advanced components of the Green's function 
are defined according to $G^{\rm 
R}(t,t')=\th(t-t')[G^{>}(t,t')-G^{<}(t,t')]$ and $G^{\rm 
A}(t,t')=-\th(t'-t)[G^{>}(t,t')-G^{<}(t,t')]$, 
and the like for the self-energy.
We emphasize that the dissipative KBE are 
exact when the correlation self-energy   
$\S^{\rm c}$ is known exactly. Otherwise, they provide an 
approximate method when  
$\S^{\rm c}[G,v]$ is treated within a diagrammatic approximation.

The correction of the HF self-energy has been included 
into a one-particle non-Hermitian Hamiltonian $h^{\rm HF}$, which is defined as 
(see Appendix~\ref{HFapp})
\begin{align}
    h^{\rm HF}_{in}(t)=h_{in}(t)+V^{\rm HF}_{in}(t)+V^{\rm a,diss}_{in}(t)-{\rm i}\g_{in}(t),
\label{meanfield h}
\end{align}
where
\begin{align}
V^{\rm HF}_{in}(t)=\sum_{mj}\left(v_{ijmn}^{\rm phys}(t)\pm 
v_{ijnm}^{\rm phys}(t)\right)\rho_{mj}^<(t),
\end{align}
is the physical (nonlocal in space) Hartree-Fock potential, and
\begin{subequations}
\begin{align}
    \g_{in}(t)&={\g}^>_{in}(t)\mp {\g}^<_{in}(t),
	\label{small gamma}\\
    \g_{in}^>(t)&=V_{in}^{\rm loss}(t)\pm\frac{1}{2}\sum_{mj} 
	v_{ijnm}^{\rm diss}(t)\rho_{mj}^>(t)
	\label{l>},\\
    \g_{in}^<(t)&=V_{in}^{\rm gain}(t)+\frac{1}{2}\sum_{mj} 
	v_{jimn}^{\rm diss}(t)\rho_{mj}^<(t)
	\label{l<},
\end{align}
\label{gammass}
\end{subequations}
is the one-particle decay-rate matrix -- in Eq.~(\ref{l>})  
$\rho^{>}(t)=\pm {\rm i} G^{>}(t,t)=\rho^{<}\pm 1$. The 
decay-rate matrices $\g^{\lessgtr}$ 
are both Hermitian and positive 
semi-definite (PSD). 
A stable steady-state solution of the
Lindblad equation is guaranteed in fermionic systems and needs 
to be checked case by case in bosonic systems. 
A simple example of unstable dynamics is provided by 
a single bosonic 
mode with a jump operator 
$\hat{L}=\hat{d}^\dagger$, which continuously pumps particles 
into the system~\cite{zhang2025directalgebraicproofnonpositivity}. 
Interestingly, the dissipative Hartree self-energy contributes 
to a purely Hermitian potential,  $V^{\rm a,diss}$, whereas the 
dissipative Fock 
self-energy renormalizes the complex potentials in ${\rm i}\g^{\lessgtr}$.

Subtracting Eq.~(\ref{kbe1}) from Eq.~(\ref{kbe2}) and using the 
definition of the retarded Green's function,  we 
have the equation of motion for the retarded component   
\begin{align}
    \left[{\rm i}\frac{\rm d}{{\rm d}t}-h^{\rm HF}(t)\right]G^{\rm R}(t,t')
    &=\d(t-t')\nn\\&+\int_{t_0}^\infty {\rm d}\bar{t}\,\Sigma^{\rm c,R}(t,\bar{t}) G^{\rm R}(\bar{t},t').
\end{align}
Likewise, we can derive the equation of motion for the advanced 
Green's function by subtracting Eq.~(\ref{kbe4}) from 
Eq.~(\ref{kbe3}). The relation $G^{\rm A}(t,t')=[G^{\rm R}(t',t)]^{\dag}$ is 
fulfilled in the dissipative case as well.  
The PSD property of $\g$  ensures the correct 
analytical property of $G^{\rm R/A}$ and, more generally, guarantees
the physical stability of the Lindbladian 
dynamics. Indeed, any effective non-Hermitian theory must ensure the 
PSD condition for the anti-Hermitian (dissipative) part of the 
Hamiltonian.         

The KBE admit a simple solution at the HF level, i.e., for 
$\S^{\rm c}=0$: 
\begin{align}\label{mf g}
    G^{\rm HF}(z,z')&=\th(t-t')T\left\{e^{-{\rm i}\int_{t'}^t {\rm 
	d}\bar{t}h^{\rm HF}(\bar{t})}\right\} G^{\rm HF}(z^{\prime *},z')\nn\\
    &+\theta(t'-t)G^{\rm HF}(z,z^*)\bar{T}\left\{e^{+{\rm i}\int_{t}^{t'} {\rm d}\bar{t}h^{\rm HF\dagger}(\bar{t})}\right\},
\end{align}
where $G^{\rm HF}(z,z^*)=G^{\rm HF,\gtrless}(t,t)$ for $z=t_{\pm}$.
The proof of Eq.~(\ref{mf g}) is provided in Appendix~\ref{HFapp}

\subsection{Conserving Approximations}
The  
equation of motion for one-particle density matrix $\rho^<(t)$ is derived by 
subtracting Eq.~(\ref{kbe3}) from Eq.~(\ref{kbe1}) and setting 
$t=t'$      
\begin{align}\label{lyapunov}
    \frac{\rm d}{{\rm d}t}\rho^<(t)=-{\rm i}h^{\rm 
	HF}(t)\rho^<(t)+{\rm i}\rho^<(t)h^{\rm HF\dagger}(t)+2{\g}^<(t)+I(t),
\end{align}
where $I(t)$ is the collision matrix 
\begin{align}\label{collision I}
    &I(t)=\nn\\
    &\pm\int_{t_0}^\infty {\rm d}\bar{t}\left[\Sigma^{\rm c,<} (t,\bar{t}) G^{\rm A}(\bar{t},t^+)+\Sigma^{\rm c,R} (t,\bar{t}) G^{<}(\bar{t},t^+)\right]+{\rm H.c.}
\end{align}
Equation (\ref{lyapunov}) reduces to a Lyapunov equation at the 
HF level, i.e., for $\Sigma^{\rm c}=0$.

We inspect the conservation properties entailed by the 
$weak$ and $strong$ global $U(1)$ symmetry. 
The equation of motion for the total 
particle number $N(t)= \Tr[ \rho^<(t)]$ reads    
\begin{align}\label{N(t)}
   \frac{\rm d}{{\rm d}t}N(t)&=\Tr [I(t)]-2\Tr \left[\left({\g}^>(t)\mp 
   {\g}^<(t)\right)\rho^<(t)-{\g}^<(t)\right]\nn\\
    &=\Tr\left[I(t)-2V^{\rm loss}(t)\rho^<(t)\pm 2 V^{\rm gain}(t)\rho^>(t)\right].
\end{align}
In general, the second and third terms are nonzero, leading to a 
non-conserved particle number -- consistent with the weak $U(1)$ symmetry 
of one-body loss and gain mechanisms. Conversely, the exact 
$I(t)$ is traceless since the particle-hole  jump operators 
possess a strong $U(1)$ symmetry. This exact property is preserved by 
all $\F$-derivable approximations to the correlation self-energy~\cite{PhysRev.124.287}:
\begin{align}\label{dFdG}
    \Sigma^{\rm c}_{mn}(z,z')=\frac{\d \F}{\d G_{nm}(z',z^+)}.
\end{align}
The functional $\F$ contains all vacuum diagrams with 
both physical and  
dissipative-induced scatterings of order higher than one. It 
is constructed  in the same 
manner as for the 
nondissipative case,  with the substitution $v^{\rm phys}(z,z')\to 
v(z,z')$. 

The proof of total number of particles conservation for 
$\F$-derivable self-energies is straightforward. 
Consider the variation of the NEGF induced by the infinitesimal $U(1)$ 
transformation    
\begin{align}\label{dG}
    \d G_{mn}(z,z')={\rm i}\left[\L(z)-\L(z')\right]G_{mn}(z,z').
\end{align}
This transformation leaves $\F$ unchanged.
Taking into account Eqs.~(\ref{dFdG}) and (\ref{dG}), 
as well as that $\S^{\rm c}$ is a nonsingular function of its contour 
times, we have
\begin{align}
    0&=\sum_{mn}\int_C {\rm d}z {\rm d}z' \Sigma^{\rm 
	c}_{mn}(z,z')\left[\L(z)-\L(z')\right]G_{nm}(z',z^+)\nn\\
    &=\int_C {\rm d}z{\rm d}z'\Tr\big[\Sigma^{\rm 
	c}(z,z')G(z',z^+)-G(z,z')\Sigma^{\rm c}(z',z^+)\big]\L(z).
\end{align}
Due to the arbitrariness of $\L(z)$, we infer that 
\begin{align}\label{trI=0}
    0&=\int_C {\rm d}z'\Tr\left[\Sigma^{\rm c}(z,z')G(z',z^+)-G(z,z')\Sigma^{\rm c}(z',z^+)\right]\nn\\
    &=\int_{t_0}^\infty {\rm d}\bar{t}\,\Tr\big[\Sigma^{\rm c,<} (t,\bar{t}) G^{\rm A}(\bar{t},t^+)+\Sigma^{\rm c,R} (t,\bar{t}) G^{<}(\bar{t},t^+)+{\rm H.c.}\big]\nn\\
    &=\pm\Tr I(t),
\end{align}

In conclusion, changes in the  total number of particles are entirely 
governed by one-body loss and gain jump operators. For these operators 
to conserve $N(t)$, the coefficients $\eta$ and $\bar{\eta}$ in 
Eqs.~(\ref{Vloss}–\ref{Vgain}) must satisfy the condition $\Tr[V^{\rm 
loss}(t)\rho^<(t) \mp V^{\rm gain}(t)\rho^>(t)] = 0$, which can only be 
fulfilled in special cases -- a relevant example is provided in 
Section~\ref{Semiconductors}.

\section{Dissipative Bethe-Salpeter equation}
\label{Dissipative BSE}

\subsection{Exact results}

The non-Hermitian Bethe-Salpeter equation (BSE) is an equation for the two-particle 
exchange-correlation (XC) function $L$, which is obtained from the 
two-particle Green's function as:  
\begin{align}
    \pm L(az_a,bz_b;cz_c,dz_d)&\equiv G_2(az_a,bz_b;cz_c,dz_d)\nn\\
    &-G_{ac}(z_a,z_c)G_{bd}(z_b,z_d).
	\label{defxcl}
\end{align}
According to the results in Section~\ref{Dissipative NEGF formalism}, 
dissipation modifies only the noninteracting Green's function $G^{0}$ 
and renormalizes the physical interaction $v^{\rm phys}(z,z')\to v(z,z')$, 
while leaving the rules for diagrammatic expansions otherwise 
unchanged. Therefore, 
the dissipative Bethe-Salpeter equation reads~\cite{svl-book_2025}
\begin{align}\label{bse l}
    &L(az_a,bz_b;cz_c,dz_d)=G_{ad}(z_a,z_d)G_{bc}(z_b,z_c)\nn\\
    &\pm\sum_{a'b'c'd'}\int_C {\rm d}z_a'{\rm d}z_b'{\rm d}z_c'{\rm d}z_d'  G_{aa'}(z_a,z_a')G_{c'c}(z_c',z_c)\nn\\
    &\times K(a'z_a',b'z_b';c'z_c',d'z_d')L(d'z_d',bz_b;b'z_b',dz_d).
\end{align}
where the irreducible kernel is related to the self-energy 
$\S=\S^{\rm s}[G,v]$ through
\begin{align}\label{k=ds/dg}
    K(az_a,bz_b;cz_c,dz_d)=\pm \frac{\d \Sigma^{\rm 
	s}_{ac}(z_a,z_c)}{\d G_{db}(z_d,z_b)}.
\end{align}

Let us define the {\it two-time} XC function as
\begin{align}
    L_{\substack{ij\\nm}}(z_1,z_2)\equiv L(iz_1,mz_2;jz_1^+,nz_2^+)=
	L_{\substack{mn\\ji}}(z_2,z_1),
	\label{2timexc}
\end{align}
which has property 
\begin{align}\label{adjoint L}
    L^*_{\substack{ij\\nm}}(z_1,z_2)=L_{\substack{ji\\mn}}(z_1^*,z_2^*)=
	L_{\substack{nm\\ij}}(z_2^*,z_1^*).
\end{align}
The two-time $L$  also emerges as a response function via the  
Kubo's formalism. It can 
be shown that the current and the density variations produced by 
$L$, which satisfies the Bethe-Salpeter equation with kernel $K=\pm \d 
\S/\d G$, can equivalently be computed from  
$G$, provided that $\S$ is $\F$-derivable, as in the case of Hermitian 
systems~\cite{svl-book_2025,attaccalite_real-time_2011}.      
Consequently, the poles of the two-time $L$ contain information about 
absorption and emission energies as well as dissipation-induced lifetimes.

The two-time XC function can also be used to screen the generalized 
interaction according to   
\begin{align}\label{w and dw}
    W_{ijmn}(z,z')=v_{ijmn}(z,z')+\d W_{ijmn}(z,z'),
\end{align}
where the fluctuation $\d W$ reads (repeated indices are summed 
over and integrals are over $\bar{z}$ and $\bar{z}'$)    
\begin{align}\label{dw}
    \d W_{ijmn}(z,z')=\pm {\rm i}\int_C  
	v_{qjmp}(z,\bar{z})L_{\substack{pq\\sr}}(\bar{z},\bar{z}')v_{isrn}(\bar{z}',z').
\end{align}
The screened interaction has the same symmetry property 
$W_{ijmn}(z,z')=W_{jinm}(z',z)$ as the bare $v$. 
Of particular relevance to the following discussion is the retarded 
component $W^{\rm R}\equiv W^{\rm T}-W^{\rm <}=W^{\rm 
>}-W^{\rm \bar{T}}$, where $W^{\rm T}(t,t')=W(t_{-},t'_{-})$ is  
time ordered  and $W^{\rm \bar{T}}(t,t')=W^{\rm 
\bar{T}}(t_{+},t'_{+})$ is  anti-time ordered. Using the explicit form of $v$ in 
Eq.~(\ref{generalized v}) and integrating over the contour times we 
find 
\begin{align}
    &W_{ijmn}^{\rm R}(t,t')=\left[v_{ijmn}^{\rm 
	phys}(t)+iv_{ijmn}^{\rm a,diss}(t)\right]\d(t-t')\pm {\rm i} \sum_{pqrs}\nn\\
    &\times\left[v_{qjmp}^{\rm phys}(t)+{\rm i}v_{qjmp}^{\rm 
	a,diss}(t)\right]L^{\rm 
	R}_{\substack{pq\\sr}}(t,t')\left[v_{isrn}^{\rm phys}(t')+{\rm i}v_{isrn}^{\rm a,diss}(t')\right]. 
\label{w r}
\end{align}
The advanced component 
$W^{\rm A}\equiv W^{\rm T}-W^{\rm >}=W^{\rm 
<}-W^{\rm \bar{T}}$ can be extracted similarly. Taking into account 
that 
\begin{align}\label{adjoint Lt}
    L^{\rm R *}_{\substack{ij\\nm}}(t_1,t_2)=-L^{\rm 
	A}_{\substack{nm\\ij}}(t_2,t_1),
\end{align}
it is straightforward to verify the relation   
$W_{ijmn}^{\rm R*}(t,t')=W_{mnij}^{\rm A}(t',t)$.

\subsection{BSE with HSEX kernel}
The lowest-order approximation to $\S$ in a skeletonic 
expansion in terms of $G$ and $W$ is given by the Hartree term in Eq.~(\ref{hartree}) 
and the Fock term in Eq.~(\ref{fock}) with $v\ra W$ (screened exchange). We 
refer to this approximation as the Hartree plus screened exchange 
(HSEX) approximation.
The HSEX kernel follows from Eq.~(\ref{k=ds/dg}).
Discarding the functional derivative of $W$ with respect to $G$, we 
find
\begin{align}\label{hsex k}
    &K^{\rm HSEX}(a'z_a',b'z_b';c'z_c',d'z_d')\nn\\
    &={\rm i}\d(z_a',z_c')\d(z_b',z_d')v_{a'b'd'c'}(z_b',z_a')\nn\\
    &\pm {\rm i}\d(z_a',z_d')\d(z_b',z_c')W_{a'b'c'd'}(z_b',z_a').
\end{align}
To evaluate the screened interaction we 
implement the (widely used) static approximation, according 
to which $W^{\rm R}(t,t')=\d(t-t')W^{\rm R}(\w=0)$. 
Physical dissipation-induced interactions are typically short-ranged 
and do not necessitate to be screened, see also Section~\ref{Valley excitons in TMD}. 
Therefore, we can safely discard $v^{\rm a,diss}$ in the second line 
of Eq.~(\ref{w r}). With these simplifications Eq.~(\ref{w and dw}) 
becomes 
\begin{align}
	W_{ijmn}(z,z')=\d(z,z')W^{\rm phys}_{ijmn}(t)+v^{\rm 
	diss}_{ijmn}(z,z'),
\label{static dw}
\end{align}
where $W^{\rm phys}=v^{\rm phys}+v^{\rm phys}L^{\rm 
R}(\w=0)v^{\rm phys}$. This screened interaction fulfills 
the same symmetry properties as 
the bare physical interaction, i.e., $ W_{ijmn}^{\rm phys *}= 
W_{nmji}^{\rm phys }=W_{mnij}^{\rm phys }$.

The main advantage of the static HSEX approximation for 
nondissipative systems is that it reduces the  
BSE to a closed form for the two-time XC 
function. As we show below, the presence of dissipation 
introduces an additional complication, which, however, can be 
effectively addressed.  
Inserting the kernel in Eq.~(\ref{hsex k}) -- with $W$ from 
Eq.~(\ref{static dw}) and $v^{\rm diss}$ from Eq.~(\ref{vdiss}) --
into the BSE Eq.~(\ref{bse l}), and integrating over $z'_{a}$, $z_c'$ and 
$z_d'$, we obtain
\begin{widetext}
\begin{align}\label{bse static hsex}
    L(az_a,bz_b;cz_c,dz_d)&=G_{ad}(z_a,z_d)G_{bc}(z_b,z_c)\pm {\rm i}\sum_{a'b'c'd'}\int_C {\rm d}z_b'\nn\\
    &\times \Big\{  G_{aa'}(z_a,z_b')G_{c'c}(z_b',z_c) \Big[v_{a'b'd'c'}^{\rm phys}(t_b')\pm W^{\rm phys}_{a'b'c'd'}(t_b')
	-{\rm i}s(z_b')\big(v_{a'b'd'c'}^{\rm s,diss}(t_b')\pm v_{a'b'c'd'}^{ \rm 
	s,diss}(t_b')\big)\Big]\nn\\
    &+  G_{aa'}(z_a,z_b'^*)G_{c'c}(z_b'^*,z_c)  \Big[{\rm i}s(z_b')v_{a'b'd'c'}^{\rm s,diss}(t_b')+{\rm i} 
	v_{a'b'd'c'}^{ \rm a,diss}(t_b')\Big]\Big\}L(d'z_b',bz_b;b'z_b',dz_d)\nn\\
    &+{\rm i}\sum_{a'b'c'd'}\int_C {\rm d}z_b'  G_{aa'}(z_a,z_b'^*)G_{c'c}(z_b',z_c) \left[{\rm i}s(z_b')v_{a'b'c'd'}^{ \rm 
	s,diss}(t_b')+{\rm i}v_{a'b'c'd'}^{ \rm 
	a,diss}(t_b')\right]L(d'z_b'^*,bz_b;b'z_b',dz_d), 
\end{align}
By setting $z_c=z_a$ and  $z_d=z_b$, all functions $L$ except the 
last one reduce to the two-time XC function in 
Eq.~(\ref{2timexc}). The origin of the last term stems from the 
anomalous Dirac delta function, $\d(z,z'^*)$, appearing in Eq.~(\ref{vdiss}).    
The crucial observation at this point is that 
\begin{align}\label{L*=L}
     L(d'z_b'^*,bz_b;b'z_b',dz_b)=L(d'z_b',bz_b;b'z_b',dz_b)
\end{align}
for all physical times  $t_b'>t_b$. Therefore, the BSE can be closed for 
the retarded two-time XC function $L^{\rm R}(t,t')$, and we find
\begin{align}\label{bse lr}
    L_{\substack{ij\\nm}}^{\rm R}(t_1,t_2)
    &=\ell_{\substack{ij\\nm}}^{\rm R}(t_1,t_2)+ {\rm i}\sum_{pqrs}\int_{t_2}^{t_1} {\rm d}\bar{t} 
	\Big\{
    \ell^{\rm R}_{\substack{ij\\pq}}(t_1,\bar{t}) 
	\Big[W_{psqr}^{\rm phys}(\bar{t}) \pm v_{psrq}^{\rm phys}(\bar{t})\mp {\rm i} v_{psrq}^{ \rm a,diss}(\bar{t})\Big]
	\nn\\&+{\rm i}\Big[G^<_{ip}(t_1,\bar{t})G_{qj}^<(\bar{t},t_1)
	-G^>_{ip}(t_1,\bar{t})G_{qj}^>(\bar{t},t_1)\Big]v_{psqr}^{ \rm a,diss}(\bar{t})
	-{\rm i}G^{\rm R}_{ip}(t_1,\bar{t})G_{qj}^{\rm 
	A}(\bar{t},t_1)v_{psqr}^{ \rm s,diss}(\bar{t})\Big\}
    L^{\rm R}_{\substack{rs\\nm}}(\bar{t},t_2),
\end{align}
\end{widetext}
where the function $\ell$ is the noncorrelated two-time XC function 
\begin{align}\label{def ell}
    \ell_{\substack{ij\\nm}}(z,z')\equiv G_{in}(z,z')G_{mj}(z',z).
\end{align}
The proof of Eq.~(\ref{L*=L}) and the derivation of Eq.~(\ref{bse 
lr}) can be found in Appendix~\ref{retBSEapp}.

\subsection{Steady-state BSE}

Let the system Hamiltonian and jump operators be independent of 
time. 
The exact numerical solution of Eq.~(\ref{bse lr}) remains an 
intractable task even for medium-sized systems, due to the many-body 
nature of the one-particle Green's function $G$.    
In practical calculations for nondissipative systems, the so-called 
quasiparticle approximation  $G\simeq G^{\rm qp}$ is commonly employed 
to simplify the BSE and obtain the 
steady-state solution. Under this approximation, the steady-state BSE 
reduces to the diagonalization of an effective two-particle 
Hamiltonian, enabling unprecedented accuracy in the description of 
optical spectra and excitonic properties of semiconductors.     

A key feature of
$G^{\rm qp}$ is the existence of well defined one-particle quantum 
numbers, which ensures well defined populations and energies, or 
equivalently, infinite quasiparticle lifetimes. This is guaranteed when 
the quasiparticle Hamiltonian $h^{\rm qp}$ commutes with the one-particle density 
matrix $\rho^{<}$ -- a condition always satisfied in the 
HF approximation since $h^{\rm qp}=h^{\rm HF}$. 

In dissipative systems the existence of well defined 
one-particle quantum numbers, necessary to construct a quasi-particle 
Green's function, is generally not guaranteed. Let us inspect the HF 
approximation.
The stationary one-particle density matrix 
satisfies the Lyapunov equation, see Eq.~(\ref{lyapunov}), 
\begin{align}
-{\rm i}h^{\rm HF}\rho^{<}+
{\rm i}\rho^{<}h^{\rm HF\dagger}+2{\g}^<=0.
\label{statlyap}
\end{align}
Thus,  the commutation condition $[h^{\rm 
HF},\rho^{<}]_{-}=0$ does not guarantee the stationarity. A 
sufficient condition for stationarity is that
the Hermitian matrices 
$(h^{\rm HF}+h^{\rm HF\dagger})$, $\g^{\gtrless}$ 
and $\rho^{<}$ can all be simulataneously 
diagonalized.
In the following we assume that this is the case, and we
evaluate  $G^{\rm qp}$  as in 
Eq.~(\ref{mf g}), replacing $h^{\rm HF}\to h^{\rm qp}$ and 
$G^{\rm HF}(t_{-},t_{+})\to \mp {\rm i} \rho^{\rm qp,<}$, where
\begin{align}
    h^{\rm qp}_{in}&=\d_{in}\left(\e_i-{\rm i}\g_i\right),\nn\\
    \rho^{\rm qp,<}_{in}&=\d_{in}f_i.
\end{align}
The corresponding quasiparticle Green's function is stationary 
provided that $\rho^{\rm qp,<}$ satisfies the Lyapunov equation with 
Hamiltonian $h^{\rm qp}$, or equivalently $f_i=\g^{<}_{i}/\g_{i}$. 
In Section~\ref{Semiconductors} we 
discuss a physical application that fulfills these conditions.   
In Appendix~\ref{HFapp} we derive all Keldysh components of $G^{\rm 
qp}$.

We evaluate Eq.~(\ref{bse lr}) using 
$G^{\rm qp}$ and then Fourier transform with 
respect to the time difference. The result is a generalization of the 
first-principles state-of-the-art Bethe-Salpeter equation 
\begin{align}\label{bse qp lr}
    &\left[\w-(\e_i-\e_j)+{\rm i}(\g_i+\g_j)\right]
	L_{\substack{ij\\nm}}^{\rm R}(\w)
    ={\rm i}\d_{in}\d_{mj}(f_j-f_i)\nn\\
    &-(f_j-f_i)\sum_{rs}\left[W_{isjr}^{\rm phys} \pm (v_{isrj}^{\rm 
	phys}- {\rm i} v_{isrj}^{ \rm a,diss})\right]
    L^{\rm R}_{\substack{rs\\nm}}(\w)\nn\\
    &+{\rm i}\sum_{rs}\left[v_{isjr}^{ \rm s,diss}\mp
	\left(f_j+f_i\pm1\right)v_{isjr}^{ \rm a,diss}\right]
	L^{\rm R}_{\substack{rs\\nm}}(\w).
\end{align}
In the second line we see that only the antisymmetric part of the 
dissipation-induced  interaction 
renormalizes the physical Hartree amplitude $v_{isrj}^{\rm phys}$. 
As for the physical Fock amplitude $W_{isjr}^{\rm phys}$, 
both the symmetric and antisymmetric parts contribute to its 
renormalization -- see  terms in the third line. Here, the 
antisymmetric contribution vanishes only for fermions at zero 
temperature since $f_j+f_i-1=0$ for all $f_j-f_i\neq 0$.
The BSE for closed system is recovered by setting 
$v^{\rm diss}=0$.

\section{Semiconductors}
\label{Semiconductors}

We specialize the general result in Eq.~(\ref{bse qp lr}) 
to fermionic systems with discrete translational symmetry and 
with a finite gap in the one-particle spectrum 
(semiconductors). If the jump operators preserve the discrete 
translational symmetry we can rewrite Eq.~(\ref{bse qp lr}) 
in a general Bloch basis  by replacing~\cite{sander_beyond_2015}
\begin{align}
    &i\ra \blk+\frac{\blq}{2} \m \quad j\ra \blk-\frac{\blq}{2} \n \quad m\ra \blk'-\frac{\blq}{2} \rho \nn\\
    &n\ra \blk'+\frac{\blq}{2} \s \quad r\ra \blk''+\frac{\blq}{2} \a \quad s \ra \blk''-\frac{\blq}{2} \b \nn
\end{align}
where $\blk$, $\blk'$, $\blk''$ and $\blq$ are quasi-momenta and greek indices 
are collective indices for band and spin. 
We introduce the short-hand notation for the two-time XC function
\begin{align}
    L_{\begin{subarray}{l}
    \m\n\blk\\\s \rho\blk'\end{subarray}}^{\blq}\equiv L_{\begin{subarray}{l}
    \blk+\frac{\blq}{2}\m\,\blk-\frac{\blq}{2}\n\\\blk'+\frac{\blq}{2}\s  \,\blk'-\frac{\blq}{2}\rho\end{subarray}}^{\blq},
\end{align}
and for the physical BSE kernel
\begin{align}
    K_{\begin{subarray}{l}\m\n\blk\\\a\b\blk''\end{subarray}}^{\blq,\rm phys}&
	\equiv W^{\rm phys}_{\blk+\frac{\blq}{2} \m\,\blk''-\frac{\blq}{2} \b\,\blk-\frac{\blq}{2} \n\,\blk''+\frac{\blq}{2} \a}
    \nn\\&-v^{\rm phys}_{\blk+\frac{\blq}{2} 
	\m\,\blk''-\frac{\blq}{2} \b\,\blk''+\frac{\blq}{2} \a\,\blk-\frac{\blq}{2} \n}
	= 
	K_{\begin{subarray}{l}\a\b\blk''\\\m\n\blk\end{subarray}}^{\blq,\rm phys\ast}.
\label{symkphys}
\end{align}
We further define
\begin{subequations}
\begin{align}
    \e_{\m\n\blk}^{\blq}&\equiv \e_{\blk+\frac{\blq}{2}\m}-\e_{\blk-\frac{\blq}{2}\n},\\
    \g_{\m\n\blk}^{\blq}&\equiv \g_{\blk+\frac{\blq}{2}\m}+\g_{\blk-\frac{\blq}{2}\n},
	\label{gammamn}\\
    f_{\m\n\blk}^{\blq}&\equiv f_{\blk-\frac{\blq}{2}\n}-f_{\blk+\frac{\blq}{2}\m}.
\end{align}
\end{subequations}
Using these definitions, the BSE in Eq.~(\ref{bse qp lr}) reads
\begin{align}\label{k-bse}
    &\left[\w-\e_{\m\n\blk}^\blq+{\rm i}\g_{\m\n\blk}^\blq\right]L_{\begin{subarray}{l}
    \m\n\blk\\\s\rho\blk'\end{subarray}}^{\blq,\rm R}(\w)= {\rm i}f_{\m\n\blk}^{\blq}\d_{\m\s}\d_{\rho\n}\d_{\blk\blk'}\nn\\
    &- 
	f_{\m\n\blk}^{\blq}\sum_{\a\b\blk''}\left(K_{\begin{subarray}{l}\m\n\blk\\\a\b\blk''\end{subarray}}^{\blq,\rm phys}+{\rm i} v^{ \rm a,diss}_{\blk+\frac{\blq}{2} \m\,\blk''-\frac{\blq}{2} \b\,\blk''+\frac{\blq}{2} \a\,\blk-\frac{\blq}{2} \n}\right)L_{\begin{subarray}{l}\a\b\blk''\\\s\rho\blk'\end{subarray}}^{\blq,\rm R}(\w)\nn\\
    &+{\rm i}\sum_{\a\b\blk''}\bigg[\left(f_{\blk-\frac{\blq}{2}\n}+f_{\blk+\frac{\blq}{2}\m}-1\right)v^{ \rm a,diss}_{\blk+\frac{\blq}{2} \m\,\blk''-\frac{\blq}{2} \b\,\blk-\frac{\blq}{2} \n\,\blk''+\frac{\blq}{2} \a}\nn\\
    &\quad\quad\quad\quad+v^{ \rm s,diss}_{\blk+\frac{\blq}{2} 
	\m\,\blk''-\frac{\blq}{2} \b\,\blk-\frac{\blq}{2} 
	\n\,\blk''+\frac{\blq}{2} \a}\bigg]L_{\begin{subarray}{l}\a\b\blk''\\\s\rho\blk'\end{subarray}}^{\blq,\rm R}(\w),
\end{align} 
Taking into account the symmetry property of $v^{\rm phys}$ and 
$v^{\rm diss}$ we verified that the exact property 
\begin{align}\label{adjoint Lw q}
    L_{\begin{subarray}{l}\m\n\blk\\\s\rho\blk'\end{subarray}}^{\blq,\rm R *}(\w)=-
    L_{\begin{subarray}{l}\s\rho\blk'\\\m\n\blk\end{subarray}}^{\blq,\rm A}(\w)
\end{align}
is satisfied.

At zero temperature and in the absence of dissipation the electronic 
populations are $f_{\blk\m}=1$ for valence bands ($\m=v$) and 
$f_{\blk\m}=0$ for conduction ($\m=c$). We consider jump 
operators that preserve this stationary (ground) state, meaning that the 
corresponding  
one-particle density matrix 
$\rho_{\blk\m\blk'\n}=\d_{\blk\blk'}\d_{\m\n}f_{\blk\m}$ satisfies the 
Lyapunov equation. A physically relevant example of such jump 
operators is discussed later. 
The BSE in Eq.~(\ref{k-bse}) can 
then be closed in the subspace of {\it resonant} and {\it antiresonant} band 
indices. A pair of indices $(\m,\n)$ is called resonant (antiresonant) if $\m$ is a 
conduction (valence) band and $\n$ is a valence (conduction) band. 
Accordingly, $f_{\m\n\blk}^{\blq}=1$ in the resonant sector and 
$f_{\m\n\blk}^{\blq}=-1$ 
in the antiresonant sector. Independently of the sector we have 
$f_{\blk-\frac{\blq}{2}\n}+f_{\blk+\frac{\blq}{2}\m}=1$ and therefore 
the third line of Eq.~(\ref{k-bse}) vanishes. It is then useful to 
introduce the dissipation kernel
\begin{align}\label{K diss}
    K_{\begin{subarray}{l}\m\n\blk\\\a\b\blk''\end{subarray}}^{\blq,\rm diss}&\equiv 
    f_{\m\n\blk}^{\blq }v^{\rm s,diss}_{\blk+\frac{\blq}{2} \m\,\blk''-\frac{\blq}{2} \b\,\blk-\frac{\blq}{2} \n\,\blk''+\frac{\blq}{2} \a}\nn\\
    &-v^{\rm a,diss}_{\blk+\frac{\blq}{2} \m\,\blk''-\frac{\blq}{2} \b\,\blk''+\frac{\blq}{2} \a\,\blk-\frac{\blq}{2} \n},
\end{align}
which shares the same symmetry property as the physical kernel, 
namely [compare with Eq.~(\ref{symkphys})]
\begin{align}\label{adjoint Kdiss}
     K_{\begin{subarray}{l}\m\n\blk\\\a\b\blk''\end{subarray}}^{\blq,\rm diss *}= 
	 K_{\begin{subarray}{l}\a\b\blk''\\\m\n\blk\end{subarray}}^{\blq,\rm diss}.
\end{align}
In deriving the above relation we use Eqs.~(\ref{dissint}) and 
(\ref{vphsa}). 
We observe  
that the prefactor $f_{\m\n\blk}^{\blq }$ in $K^{\rm diss}$ flips  the 
sign of the first term when going from the 
resonant to the antiresonant
sector.  Time-reversal symmetry is thus broken. 
It is also worth noticing that 
the symmetric interaction $v^{\rm s,diss}$ renormalizes  the Fock 
channel only while 
the antisymmetric interaction $v^{\rm a,diss}$ renormalizes  the 
Hartree channel only. 

At zero temperature the BSE in Eq.~(\ref{k-bse}) takes the following 
compact form
\begin{align}\label{k-bse in S}
    &\left[\w-\e_{\m\n\blk}^\blq+{\rm i}\g_{\m\n\blk}^\blq\right]L_{\begin{subarray}{l}
    \m\n\blk\\\s\rho\blk'\end{subarray}}^{\blq,\rm R}(\w)= {\rm i}f_{\m\n\blk}^{\blq}\d_{\m\s}\d_{\rho\n}\d_{\blk\blk'}\nn\\
    &- f_{\m\n\blk}^{\blq}\sum_{\a\b\blk''}\left(K_{\begin{subarray}{l}\m\n\blk\\\a\b\blk''\end{subarray}}^{\blq,\rm phys}-
	{\rm i} 
	K_{\begin{subarray}{l}\m\n\blk\\\a\b\blk''\end{subarray}}^{\blq,\rm diss}\right)L_{\begin{subarray}{l}\a\b\blk''\\\s\rho\blk'\end{subarray}}^{\blq,\rm R}(\w).
\end{align} 
The resonant and antiresonant 
sectors are coupled by the physical kernel 
$K^{\blq,\rm phys}$ and by the dissipation kernel $K^{\blq,\rm diss}$.
In most first-principles calculation the physical coupling is discarded, a 
scheme known as the Tamm-Dancoff approximation (TDA)~\cite{hirata_time-dependent_1999}. 
The possibility of implementing 
the TDA  for $K^{\blq,\rm diss}$ must be carefully assessed on a 
case-by-case basis. The dissipative 
resonant-antiresonant coupling reads
\begin{align}
    K_{\begin{subarray}{l}cv\blk\\v'c'\blk'\end{subarray}}^{\blq,\rm diss}
    &=v^{\rm s,diss}_{\blk+\frac{\blq}{2} c\,\blk'-\frac{\blq}{2} c'\,\blk-\frac{\blq}{2} v\,\blk'+\frac{\blq}{2} v'}
	-v^{\rm a,diss}_{\blk+\frac{\blq}{2} c\,\blk'-\frac{\blq}{2} 
	c'\,\blk'+\frac{\blq}{2} v'\,\blk-\frac{\blq}{2} v},
	\label{rarc}
\end{align}
while the antiresonant-resonant coupling follows from Eq.~(\ref{adjoint 
Kdiss}). 

\subsection{Stationarity-Preserving Jump Operators}
We consider the non-Hermitian jump operators preserving the 
discrete translational symmetry of the underlying lattice [compare 
with Eq.~(\ref{genlind})]
\begin{subequations}
\begin{align}
    \hat{L}_{\blk c}^{\rm p}&= \eta^{\blk c}\hat{d}_{\blk c},\label{loss app}\\
    \hat{L}_{\blk v}^{\rm h}&=\bar{\eta}^{\blk v *}\hat{d}^\dagger_{\blk v},\label{gain app}\\
    \hat{L}_{\blq}^{\rm ph}&=\sum_{\blk \m} \c_{\blk \m\m}^{\blq}
	\hat{d}_{\blk-\frac{\blq}{2} \m}^\dagger \hat{d}_{\blk+\frac{\blq}{2}\m}
    +\sum_{\blk vc}\c^{\blq}_{\blk  vc} \hat{d}_{\blk-\frac{\blq}{2} v}^\dagger 
	\hat{d}_{\blk+\frac{\blq}{2}c}.\label{phloss app}
\end{align}
\end{subequations}
The  operators in Eqs.~(\ref{loss app}-\ref{gain app})
describe one-particle loss from conduction bands and 
one-particle gain from valence bands. The operator in 
Eq.~(\ref{phloss app}) is the sum 
of two terms. The first is responsible for scattering particles 
within the same band whereas the second is responsible for  
particle-hole recombination. Using Eq.~(\ref{dissint}), the 
dissipation-induced interaction arising from $\hat{L}_{\blq}^{\rm 
ph}$ reads
\begin{align}\label{vphapp}
    v^{\rm diss}_{\blk_1\m_1\,\blk_2\m_2\,\blk_3\m_3\,\blk_4\m_4}
	&=2\d_{\blk_1-\blk_4,\,\blk_3-\blk_2}
	\nn\\
	&\times \c^{\blk_1-\blk_4 *}_{\frac{\blk_1+\blk_4}{2} \m_4\m_1}\c^{\blk_3-\blk_2}_{\frac{\blk_2+\blk_3}{2}\m_2\m_3}
	F_{\m_{1}\m_{2}\m_{3}\m_{4}}
\end{align}
where the tensor $F_{cvc'v'}=F_{\m\n\n\m}=F_{\m vc\m}=F_{c\m\m v}=1$, and 
vanishes otherwise.

Let us prove that the jump operators in Eqs.~(\ref{loss 
app}-\ref{phloss app}) preserve the ground state with 
populations $f_{\blk v}=1-f_{\blk c}=1$.
The strategy is as follows: we calculate $h^{\rm HF}$ in Eq.~(\ref{meanfield h}) and 
$\g^{\lessgtr}$ in Eqs.~(\ref{gammass}) 
using the ground state density matrix 
$\rho^{<}$ of the nondissipative system, and 
then we verify that the same $\rho^{<}$ satisfies the stationary Lyapunov 
equation in Eq.~(\ref{statlyap}). Due to the discrete translational 
invariance all quantities are diagonal in momentum 
space. In particular, the ground state density matrix  has diagonal elements 
$\rho^{<}_{\blk\m\n}=\d_{\m\n}f_{\blk\m}$.
By construction the commutator 
$[h+V^{\rm HF},\rho^{<}]_{-}=0$ (stationarity of the nondissipative 
system). It is easy to show that the potential 
$V^{\rm a,diss}$ and rates $\g^{\lessgtr}$ are diagonal in the 
band-indices, and that the diagonal elements read
\begin{align}
V^{\rm a,diss}_{\blk\m}&=0	\label{R app}\\
    {\g}^>_{\blk 
	\m}&=\d_{\m\{c\}}\left[|\eta^{\blk\m}|^2+\sum_{\blk'}\Big|\c^{\blk-\blk' }_{\frac{\blk+\blk'}{2}\m\m}\Big|^2\right],\label{l> app}\\
    {\g}^<_{\blk 
	\m}&=\d_{\m\{v\}}\left[|\bar{\eta}^{\blk\m}|^2+\sum_{\blk'}\Big|\c^{\blk'-\blk }_{\frac{\blk'+\blk}{2}\m\m}\Big|^2\right],\label{l< app}
\end{align}
In Eqs.~(\ref{l> app}-\ref{l< app}) the quantity $\d_{\m\{c\}}=1$ 
($\d_{\m\{v\}}=1$) if 
$\m$ is a conduction (valence) band-index and zero otherwise. Thus, 
the 
stationary Lyapunov equation is solved by the one-particle density 
matrix
\begin{align}
    \rho^<_{\blk\m\n}=\d_{\m\n}\frac{{\g}^<_{\blk\m}}{{\g}_{\blk\m}}=\d_{\m\n}\frac{{\g}^<_{\blk\m}}{{\g}^>_{\blk\m}+{\g}^<_{\blk\m}}=\d_{\m\n}f_{\blk\m},
\end{align}
which proves our original statement. We emphasize that excitations over 
the ground state still manifest nontrivial dissipative 
behavior, as we show in Section~\ref{Valley excitons in TMD}.        

The stationarity-preserving jump operators in Eq.~(\ref{phloss 
app}) have another remarkable feature: the TDA for the dissipation 
kernel is not an approximation, but an exact property. The 
resonant-antiresonant coupling is governed by a dissipation-induced 
interaction $v^{\rm diss}$ with band-indices $(c,c',v,v')$, see 
Eq.~(\ref{rarc}). However, for this specific choice of band-indices 
$v^{\rm diss}=0$ since $F_{cc'vv'}=0$, see discussion below 
Eq.~(\ref{vphapp}). This implies that we can solve the BSE in the 
resonant and antiresonant sectors separately.

\subsection{Non-Hermitian excitonic Hamiltonian}
\label{nonHsec}

In the resonant subspace, the BSE in Eq.~(\ref{k-bse in S}) reduces to 
the following eigenvalue problem
\begin{align}\label{resonant bse R}
    &\sum_{c''v''\blk''}\left(\d_{cc''}\d_{vv''}\d_{\blk\blk''}\w-
    H^\blq_{cv\blk\,c''v''\blk''}\right)
    L_{\begin{subarray}{l}c''v''\blk''\\c'v'\blk'\end{subarray}}^{\blq,\rm R}(\w)\nn\\
    &= {\rm i}\d_{cc'}\d_{vv'}\d_{\blk\blk'}
\end{align} 
where the non-Hermitian excitonic Hamiltonian
\begin{align}\label{Hq}
    H^\blq_{cv\blk\,c'v'\blk'}=H^{\blq,\rm phys}_{cv\blk\,c'v'\blk'}-{\rm i}\G^\blq_{cv\blk\,c'v'\blk'},
\end{align}
comprises the Hermitian excitonic Hamiltonian of the nondissipative system
\begin{align}
    H^{\blq,\rm 
	phys}_{cv\blk\,c'v'\blk'}=&\d_{cc'}\d_{vv'}\d_{\blk\blk'}\e_{cv\blk}^\blq- K_{\begin{subarray}{l}cv\blk\\c'v'\blk'\end{subarray}}^{\blq,\rm phys},\label{H coul}
\end{align}
and the Hermitian dissipation-induced rate 
\begin{align}
    \G^{\blq}_{cv\blk\,c'v'\blk'}=&\d_{cc'}\d_{vv'}\d_{\blk\blk'}\g_{cv\blk}^\blq- 
	K_{\begin{subarray}{l}cv\blk\\c'v'\blk'\end{subarray}}^{\blq,\rm diss},
	\label{Gamma}
\end{align}
The non-Hermitian excitonic Hamiltonian in the antiresonant sector can
be obtained similarly.

It is worth remarking that the microscopic diagrammatic construction 
of the BSE does not guarantee that $\G^{\blq}$ is PSD for any choice of the 
coefficients $\c$ in the particle-hole jump operators. 
The PSD property is crucial for preserving the causality of the
retarded XC function, and it should be satisfied by any sensible
approximation. Below we prove that $\G^{\blq}$ in Eq.~(\ref{Gamma}) is 
PSD only provided that the band-diagonal coefficients 
satisfy the relation $\c^{\blq}_{\blk\m\m}=\c^{-\blq}_{\blk\m\m}$ or 
$\c^{\blq}_{\blk\m\m}=-\c^{-\blq}_{\blk\m\m}$. 

According to Eq.~(\ref{gammamn}) we have 
$\g_{cv\blk}^\blq=\g_{\blk+\frac{\blq}{2}c}+\g_{\blk-\frac{\blq}{2}v}$, 
where we recall that $\g_{\blk\m}=\g^{>}_{\blk\m}+\g^{<}_{\blk\m}$, 
see Eq.~(\ref{small gamma}), with $\g^{\lessgtr}_{\blk\m}$ given 
in Eqs.~(\ref{l> app}-\ref{l< app}). To evaluate the dissipation kernel 
we use Eq.~(\ref{vphapp}),  
extract the symmetric $v^{\rm s,diss}$ and antisymmetric $v^{\rm 
a,diss}$ interactions, and then insert the results into 
Eq.~(\ref{rarc}). We find
\begin{align}\label{Gamma app}
    \G^\blq_{cv\blk\,c'v'\blk'}&=\d_{cc'}\d_{vv'}\d_{\blk\blk'}\bigg(\big|\eta^{\blk+\frac{\blq}{2} c}\big|^2+
	\big|\bar{\eta}^{\blk-\frac{\blq}{2} v}\big|^2\nn\\
    &+\sum_{\blp}\Big|\c^{\blk-\blp }_{\frac{\blk+\blp+\blq}{2}cc}\Big|^2+
	\sum_{\blp}\Big|\c^{\blp-\blk}_{\frac{\blk+\blp-\blq}{2}vv}\Big|^2
    \bigg)\nn\\
    &-\d_{cc'}\d_{vv'}\bigg(\c^{\blk-\blk' 
	*}_{\frac{\blk+\blk'+\blq}{2}cc}\c^{\blk-\blk' }_{\frac{\blk+\blk'-\blq}{2}vv} 
    +\c^{\blk'-\blk 
	}_{\frac{\blk'+\blk+\blq}{2}cc}\c^{\blk'-\blk *}_{\frac{\blk+\blk'-\blq}{2}vv} \bigg)
	\nn\\
	&+\c^{\blq *}_{\blk vc}\c^{\blq}_{\blk'v'c'}
\end{align}
The terms in the first two lines arise from $\g$, the third line 
arises from $v^{\rm s,diss}$, and the fourth line arises from $v^{\rm 
a,diss}$. The contributions to $\G^{\blq}$ from the first and fourth 
lines are PSD by inspection. The contributions from the second and third 
lines are diagonal in the band indices.  If we define 
$C_{\blk\blk'}\equiv \c^{\blk-\blk'}_{\frac{\blk+\blk'+\blq}{2}cc}$ and 
$V_{\blk\blk'}\equiv \c^{\blk'-\blk}_{\frac{\blk+\blk'-\blq}{2}vv}$, 
then we are left to prove that the matrix 
\begin{align}
M_{\blk\blk'}=\d_{\blk\blk'}\sum_{\blp}\big(|C_{\blk\blp}|^{2}+|V_{\blk\blp}|^{2}\big)
-C^{\ast}_{\blk\blk'}V_{\blk'\blk}-C_{\blk'\blk}V^{\ast}_{\blk\blk'}
\end{align}
is PSD. At this point of the 
derivation we have not yet used that $\c^{\blq}_{\blk\m\m}$ is even 
or odd in $\blq$. If it is, then both $C$ and $V$ are symmetric or 
antisymmetric matrices, 
and the diagonal elements of $M$ satisfy the inequality
\begin{align}
M_{\blk\blk}&=\sum_{\blp\neq \blk}\big(|C_{\blk\blp}|^{2}+|V_{\blk\blp}|^{2}\big)
+|C_{\blk\blk}-V_{\blk\blk}|^{2}
\nn\\
&\geq  \sum_{\blp\neq \blk}\big(|C_{\blk\blp}|^{2}+|V_{\blk\blp}|^{2}\big)
\geq \sum_{\blp\neq \blk}\big|
C^{\ast}_{\blk\blp}V_{\blp\blk}+
C_{\blp\blk}V_{\blk\blp}^{\ast}\big|\geq 0.
\label{inM}
\end{align}
According to Gershgorin theorem, for every eigenvalue $\l$ of the 
matrix $M$ there exists a $\blk=\blk_{\l}$ such that
\begin{align}
|\l-M_{\blk_{\l}\blk_{\l}}|&\leq \sum_{\blp\neq 
\blk_{\l}}|M_{\blk_{\l}\blp}|=\sum_{\blp\neq 
\blk_{\l}}|C^{\ast}_{\blk_{\l}\blp}V_{\blp\blk_{\l}}
+C_{\blp\blk_{\l}}V^{\ast}_{\blk_{\l}\blp}|
\nn\\
&\leq M_{\blk_{\l}\blk_{\l}}
\end{align}
where we use Eq.~(\ref{inM}) in the last inequality. We conclude 
that $\l\in (0,2M_{\blk_{\l}\blk_{\l}})$ and hence that $\l\geq 0$.

In a similar manner, one can prove that $\G^\blq_{vc\blk\,v'c'\blk'}$ 
(antiresonant sector) is PSD, since it has the same form as 
Eq.~(\ref{Gamma app}), with the replacements $\blq\to-\blq$ and 
$\c\to\c^{\ast}$.
It is worth commenting on the inclusion of the jump operator $\sum_{\blk cv}
\c^{\blq}_{cv\blk}\hat{d}^{\dag}_{\blk-\frac{\blq}{2}c}
\hat{d}_{\blk+\frac{\blq}{2}v}$ to Eq.~(\ref{phloss app}). 
This jump operator 
promotes an electron from the 
valence band to the conduction band and is often used to model a 
pump~\cite{munozarboleda2025thermodynamicsanalogueblackholes,wang2024non}.
However, its inclusion would not only destabilize the ground state, 
but also jeopardize the PSD property of $\G^{\blq}$,
as it modifies Eq.~(\ref{Gamma app}) by adding the matrix 
$-\c^{\blq\ast}_{\blk cv} \c^{\blq}_{\blk'c'v'}$, which is manifestly negative 
definite.

\section{Valley excitons in TMD}
\label{Valley excitons in TMD}

In this section we apply our theory to a two-dimensional transition 
metal dichalcogenide (TMD) system, which has band extrema located at 
the triangular lattice Brillouin zone corners $K$ and $K'$.

Let us write the physical (Coulomb) kernel in 
Eq.~(\ref{symkphys}) as 
$K^{\rm phys}=K^{W}+K^{v}$ where $K^{W}$ is the direct screened 
interaction (first term) and $K^{v}$ is the bare exchange 
interaction (second term).
The two valley excitons are 
degenerate if we set $K^{v}=0$. We define 
our unpertubed 
excitonic basis as the eigenvectors of the eigenvalue problem
\begin{align}
    \sum_{c'v'\blk'}
	\big(
\d_{cc'}\d_{vv'}\d_{\blk\blk'}\e_{cv\blk}^\blq- 
K_{\begin{subarray}{l}cv\blk\\c'v'\blk'\end{subarray}}^{\blq,W}
\big)A_{c'v'\blk'}^{\t,\blq}=E_{q} A_{cv\blk}^{ \t,\blq},
\end{align}
where $E_{q}\simeq E_{0}+\frac{\hbar^2q^2}{2M_0}$ is the degenerate exciton 
energy, and $\t=+$ for the $K$ exciton and $\t=-$ for the $K'$ 
exciton. We take $E_{0}\simeq 2$~eV as the reference energy, and the 
excitonic mass $M_0$ approximately equal to the free electron 
mass~\cite{RevModPhys.90.021001}. 

In the chosen basis the nondissipative excitonic Hamiltonian
reduces to the following $2\times 2$ 
matrix~\cite{yu2014dirac,yu2015valley,qiu_nonanaliticity_2015,wu2015exciton,wang2024non} 
\begin{align}
    &H^{\blq,\rm 
	phys}_{\t\t'}=\sum_{\substack{cv\blk\\c'v'\blk'}}A_{cv\blk}^{ 
	\t,\blq 
	*}H^{\blq,\rm phys}_{cv\blk\,c'v'\blk'}A_{c'v'\blk'}^{ \t',\blq}\nn\\
    &=\left(\frac{\hbar^2q^2}{2M_0}+U_q\right)\d_{\t\t'}+\left(\begin{array}{cc}
0 & U_q e^{-2 i \vf} \\
U_qe^{2 i \vf} & 0
\end{array}\right)_{\t\t'},
\label{phys22}
\end{align}
where $\blq = q\left(\cos \vf,\sin \vf\right)^{\top}$ is expressed in polar 
coordinates, and the bare exchange 
interaction    (stemming from $K^{v}$) 
\begin{align}\label{Jq}
    U_q=U\frac{q}{q_{K}}
\end{align}
scales linearly with $q$ in the long-wavelength 
limit~\cite{wu2015exciton,yu2014dirac}. In Eq.~(\ref{Jq}) the 
quasimomentum $q_{K}=4\pi/3a$ is the distance between the $\G$ 
point and the $K$, $K'$ points, $a\approx 3.3 \text{\AA}$ being a 
typical lattice constant of 
TMD monolayers~\cite{PhysRevB.84.153402}.    
The value $U\sim 1$ eV has been extracted 
from first principles calculations~\cite{yu2014dirac}.
Quasimomenta in the light cone satisfy the approximate inequality 
$q<q_{c}=E_{0}/(\hbar c)\sim 10^{-3}q_{K}$. 

We study the TMD coupled to photon baths.
In this context, the relevant jump 
operators describe band off-diagonal particle-hole recombination 
processes.
Therefore, we set to zero the loss and gain coefficients, 
$\eta^{\blk\m}=\bar{\eta}^{\blk\m}=0$, along with the band 
diagonal coefficients, $\c^{\blq}_{\blk\m\m}=0$.
Accordingly, see 
Eq.~(\ref{Gamma app}),
$\G_{cv\blk\, c'v'\blk'}^\blq=\c_{\blk vc}^{\blq 
*}\c_{\blk' v'c'}^\blq$, where the band off-diagonal $\c$-coefficients have the form
\begin{align}
\c^{\blq}_{\blk 
vc}=\sqrt{D_{q}}\,\blE(\blq)
\cdot\bld^{\blq}_{\blk vc}.
\end{align}
In this equation $\bld^{\blq}_{\blk vc}$ is the unit vector of dipole matrix elements, 
$\blE(\blq)$ is the unit vector of the electric field generated by 
photons of momentum $\blq$, and $D_{q}\geq 0$ is a $q$-dependent proportionality constant. 
The $2\times 2$ excitonic Hamiltonian in Eq.~(\ref{phys22}) is 
modified by the addition of the $2\times 2$ matrix $-{\rm 
i}\G^{\blq}$, where
\begin{align}
    \G^{\blq}_{\t\t'}
    &=\sum_{\substack{cv\blk\\c'v'\blk'}} 
	A_{cv\blk}^{\t,\blq *}\G_{cv\blk\, c'v'\blk'}^\blq 
	A_{c'v'\blk'}^{\t',\blq}=\c_{\t}^{\blq 
*}\c_{\t'}^\blq.
\end{align}
By construction, the $\c$-coefficients in the excitonic basis are given by
\begin{align}
\c_{\t}^\blq\equiv \sum_{cv\blk} A_{cv\blk}^{\t,\blq}
\c_{\blk vc}^\blq=\sqrt{D_{q}}\,\blE(\blq)\cdot \bld^{\blq}_{\t}
\label{cinxbasis}
\end{align}
where $\bld^{\blq}_{\t}\equiv\sum_{cv\blk}A^{\t,\blq}_{cv\blk}\bld^{\blq}_{\blk 
vc}$ is the dipole moment for the recombination of the unperturbed exciton 
in valley $\t$. For small momentum transfer $\bld^{\blq}_{\t}= 
(1,\t {\rm i})^{\top}/\sqrt{2}$, which is independent of $\blq$
~\cite{lin_essential_2023,peng_tailoring_2022}.

We consider two distinct baths that are engineered to support structured photons.
 The first 
bath (B0) is made of linearly polarized photons with no net orbital angular 
momentum (OAM); hence $\blE(\blq)=(\cos\th_{0},\sin\th_{0})^{\top}$, where the 
angle $\theta_{0}$ defines the polarization direction. The second 
bath (B1) is more exotic as it is made of two types of photons: 
type ``$+$'' photons, which are right-handed 
circularly polarized and carry an OAM $l=1$, 
and type ``$-$'' photons, which have opposite 
handedness and OAM. 
In B1 the electric field reads 
$\blE(\blq)=\blE_{+}(\blq)+\blE_{-}(\blq)$, where 
$\blE_{\pm}(\blq)= e^{\pm {\rm i}(\vf+\th_{1})}(1,\mp{\rm i})^{\top}/\sqrt{2}$
is the electric field generated by the type $\pm$ photons.
The phase $\th_{1}$ is a free parameter which, as we see below, 
controls the symmetry breaking and 
the existence of EPs. 

To our knowledge, experimental realizations of structured optical 
cavities with smooth spectral densities designed to mimic such baths 
have not yet been realized. However, significant progress has 
been made in confining photons with both circular polarization and 
finite OAM, particularly through the use 
of spiral-phase mirrors and chiral metasurfaces~\cite{sam-oam_conv,quantum1010010,marrucci2011spin,yang_topological_2022}.
We also note that structured photon baths can naturally emerge from a 
more refined treatment of the system–bath 
decoupling~\cite{jager_lindblad_2022}. In any case, our results 
depend solely on the form of the $\c$-coefficients -- the 
photon baths serve to exemplify a possible microscopic 
realization. 

Let us now discuss the modeling of the function $D_{q}$.
Since electron-hole recombination mainly occurs for momenta 
within the light cone, we let the magnitude $D_q$ vanish 
exponentially as $q\to \iif$. The small $q$ behavior of $D_{q}$ 
depends on the nature of the bath.
For bath B0, the electric field in real 
space is a plane wave, and therefore we can take $D_q\to D^{(0)}$ as $q\to 0$. 
Electric fields with OAM $l$, such as those of bath B1,
are typically generated using Laguerre-Gaussian beams.  
In real space they have the form 
$E(r,\vf_{r})\propto r^{|l|}e^{-r^{2}}e^{il\vf_{r}}$, where 
$\blr=r(\cos\vf_{r},\sin\vf_{r})^{\top}$, see 
Refs.~\cite{yao_orbital_2011,shen_optical_2019,otte_optical_2020,emile_OAM_2025}. 
This implies that, in momentum space, $E(q,\vf)\sim q^{|l|}e^{il\vf}$ as 
$q\to 0$. Therefore, we can take $D_q\to  D^{(1)}\frac{q}{q_{K}}$ in the 
same limit. To summarize
\begin{align}\label{Dq}
    D_q=\left\{\begin{array}{lr}
	D^{(0)}_{q}=D^{(0)}e^{-\l q}& {\rm (B0)}
\\
    D^{(1)}_{q}=D^{(1)}\frac{q}{q_{K}}e^{-\l q}&{\rm (B1)}
	\end{array}\right.,
\end{align}
where $D^{(0)},D^{(1)}\geq 0$ and $\l\gtrsim 1/q_{c}$.

From Eq.~(\ref{cinxbasis}) we have
\begin{align}
\c_{\t}^\blq=\sqrt{D^{(0)}_{q}}\,e^{\iu\t\th_{0}}\quad{\rm (B0)};
\quad
\c_{\t}^\blq=\sqrt{D^{(1)}_{q}}\,e^{\iu\t(\vf+\th_{1})}\quad{\rm (B1)}.
\end{align}
The two different baths can be treated simulataneously if we introduce the 
parameter $n$ with value $0$ for bath B0 and value $1$ for bath B1:
$\c_{\t}^\blq=\sqrt{D^{(n)}_{q}}\,e^{\iu\t(n\vf+\th_{n})}$. The explicit form 
of the $2\times 2$ matrix $\G^{\blq}$ reads
\begin{align}
    \G^{\blq}_{\t\t'}
    =\left(\begin{array}{cc}
    D^{(n)}_q &
    D^{(n)}_q e^{-2{\rm i}(n\vf+\th_{n})} \\
    D^{(n)}_q e^{+2{\rm i}(n\vf+\th_{n})}&
    D^{(n)}_q
    \end{array}\right)_{\t\t'}.
\end{align}

In the $2\times 2$ subspace the non-Hermitian excitonic Hamiltonian
is $H^{\blq}_{\t\t'}=H^{\blq,\rm phys}_{\t\t'}-{\rm 
i}\G^{\blq}_{\t\t'}$, see Eq.~(\ref{Hq}),
and  the 
left  and right eigenvalue problems read
\begin{align}
    H^\blq  \vec{u}_{\blq,\rm R}^\pm &= E_\blq^\pm \vec{u}_{\blq,\rm R}^\pm,\nn\\
    H^{\blq \dagger} \vec{u}_{\blq,\rm L}^\pm &= E_{\blq}^{\pm *} \vec{u}_{\blq,\rm L}^\pm .
\end{align}
We find eigenvalues
\begin{align}\label{eigenvalue}
    E_\blq^\pm&=\frac{\hbar^2q^2}{2M_0}+U_q-{\rm i} D^{(n)}_q \nn\\
    &\pm\sqrt{U_q^2-D^{(n)2}_q-2{\rm i}U_qD^{(n)}_q\cos[2\th_{n}-2(1-n)\vf]}
\end{align}
and unnormalized right/left eigenvectors 
\begin{subequations}
\begin{align}
    \vec{u}_{\blq,\rm R}^\pm &=\left(\begin{array}{c}
    \pm  {e^{- {\rm i} \vf}}\sqrt{U_q-{\rm i}e^{-2i[(n-1)\vf+\th_{n}]}D^{(n)}_q}\\
    e^{ {\rm i} \vf}\sqrt{U_q-{\rm i}e^{2{\rm i}[(n-1)\vf+\th_{n}]}D^{(n)}_q}
    \end{array}\right)
	\label{eigenvectorR}\\
    \vec{u}_{\blq,\rm L}^\pm&=\left(\begin{array}{c}
    \pm  {e^{- {\rm i} \vf}}\sqrt{U_q+{\rm i}e^{-2{\rm 
	i}[(n-1)\vf+\th_{n}]}D^{(n)}_q}\\
    e^{ {\rm i} \vf}\sqrt{U_q+{\rm i}e^{2{\rm i}[(n-1)\vf+\th_{n}]}D^{(n)}_q}
    \end{array}\right).
\end{align}
\label{eigenvector}
\end{subequations}
Note that the left and right eigenvectors can be converted into each 
other by shifting  $\th_{n}\ra\th_{n} +\frac{\pi}{2}$.

\subsection{Exceptional points}

The angular dependence of the excitonic eigenvalues for the two baths 
differs solely in the argument of the cosine, see Eq.~(\ref{eigenvalue}). 
In B0  we have $\cos[2(\th_{0}-\vf)]$ 
whereas in B1 we have $\cos(2\th_{1})$. Therefore, B0 introduces an 
anisotropy in momentum space, while B1 preserves cylindrical symmetry 
within each valley. We begin by discussing B1.   

\begin{figure}[tbp]
\includegraphics[width=0.49\textwidth]{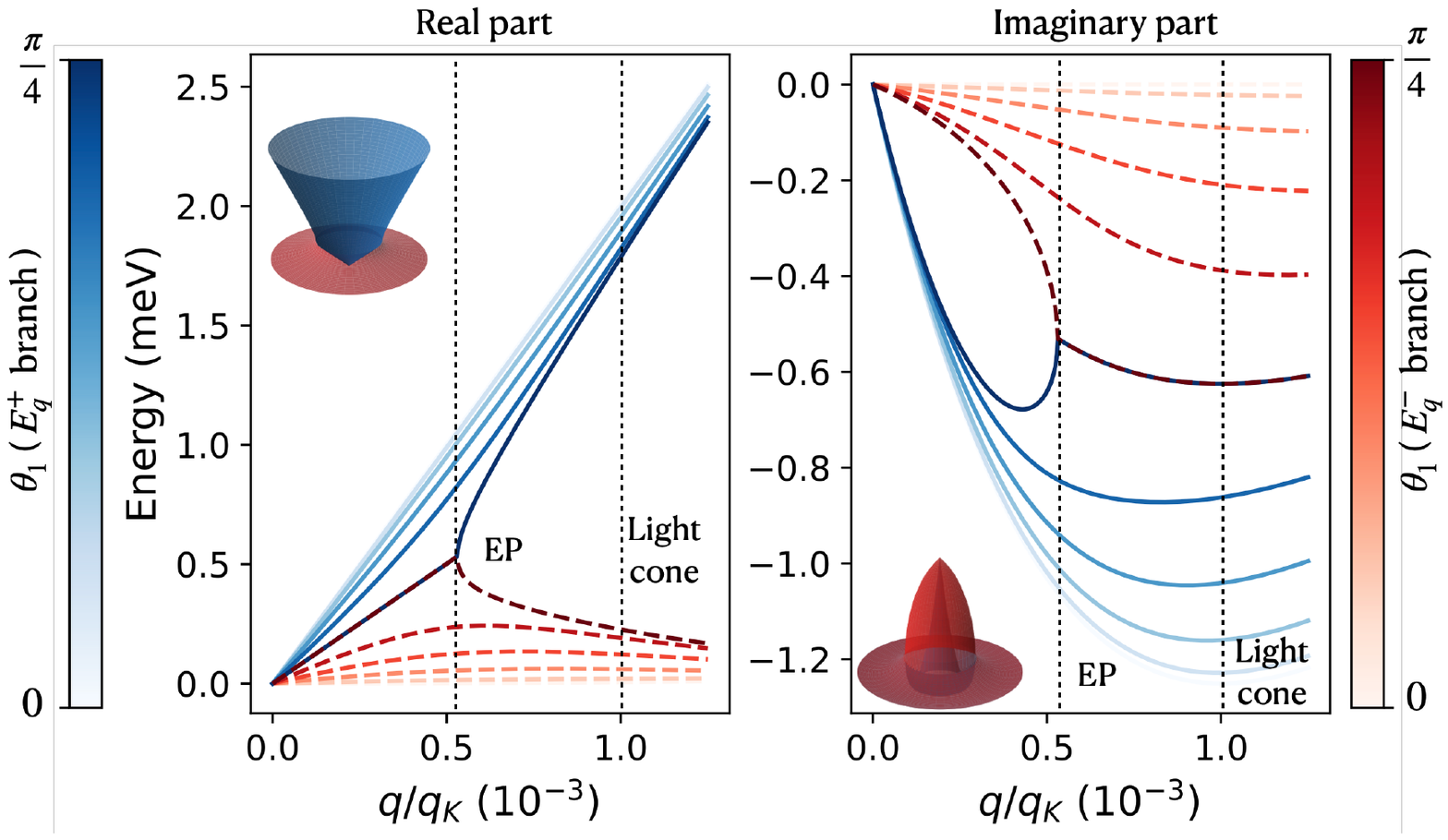}
\caption{\label{fig:ReE} Excitonic spectrum for bath B1.
Real (left) and imaginary (right) 
part of the  eigenvalues 
$E^\pm_\blq$ as a function of the exciton momentum $q$ for 
different symmetry-breaking parameters $\theta_{1}$. The upper and lower 
branches are shown in solid-blue and dashed-red, respectively. At 
$\theta_{1}=\pi/4$, 
a ring of EP emerges inside the light cone. The insets 
show the three-dimensional dispersion at 
$\theta_{1}=\pi/4$, highlighting the cylindrical symmetry.}                 
\end{figure}

In Fig.~\ref{fig:ReE} we plot the 
real and imaginary parts of the eigenvalues $E_\blq^\pm$ as 
functions of the exciton momentum $q$, 
for various values of the parameter $\th_{1}$. 
The dissipation parameters are $D^{(1)}=1.7 U$ and $\l=1/q_{c}$.
The imaginary parts are both negative, in agreement with the 
general PSD proof we derived in Section~\ref{nonHsec}.
The eigenvalues 
are well separated for all $q$, except at $\theta_{1}=\pm \pi/4$, where 
both their real and imaginary parts 
coalesce at the exceptional momentum
$q\sim q_{e}$ (inside the light cone).
The momentum $q_{e}$ follows from Eq.~(\ref{eigenvalue}) with 
$\th_{1} =\pm\frac{\pi}{4}$. Setting the square root to zero we find 
$q_{e}=\frac{1}{\l}\ln\frac{D^{(1)}}{U}\simeq 0.53q_{c}$. 
Notice that exceptional momenta are found only for $D^{(1)}>U$.
At $q=q_{e}$
one of the off-diagonal elements of the non-Hermitian excitonic 
Hamiltonian $H^{\blq}$ becomes zero, resembling an effective 
nonreciprocal coupling as encountered in quantum 
optics~\cite{PhysRevA.111.033528}.
The degenerate eigenvalues take the form
\begin{align}
    E_{\blq_{e}}^\pm=
	\frac{\hbar^2q_e^2}{2M_0}+\frac{q_{e}}{q_{K}}(U-iD^{(1)}
	e^{-\l q_e}),
\end{align}
and for, e.g., $\th_{1}=\p/4$ 
the right/left eigenvectors (normalized to unity)
\begin{align}
    \vec{u}_{\blq_{e},\rm R}^\pm =\left(\begin{array}{c}
    0\\
    1
    \end{array}\right),\quad 
	\vec{u}_{\blq_{e},\rm R}^\pm &=\left(\begin{array}{c}
    1\\
    0
    \end{array}\right),
\end{align}
coalesce. 
As the eigenvalues $E_{\blq}^\pm$ are independent of the angular coordinate 
$\vf$ of $\blq$,  a ring of exceptional points (EP) emerges 
at momentum $q=q_{e}$.

\begin{figure}[tbp]
\includegraphics[width=0.49\textwidth]{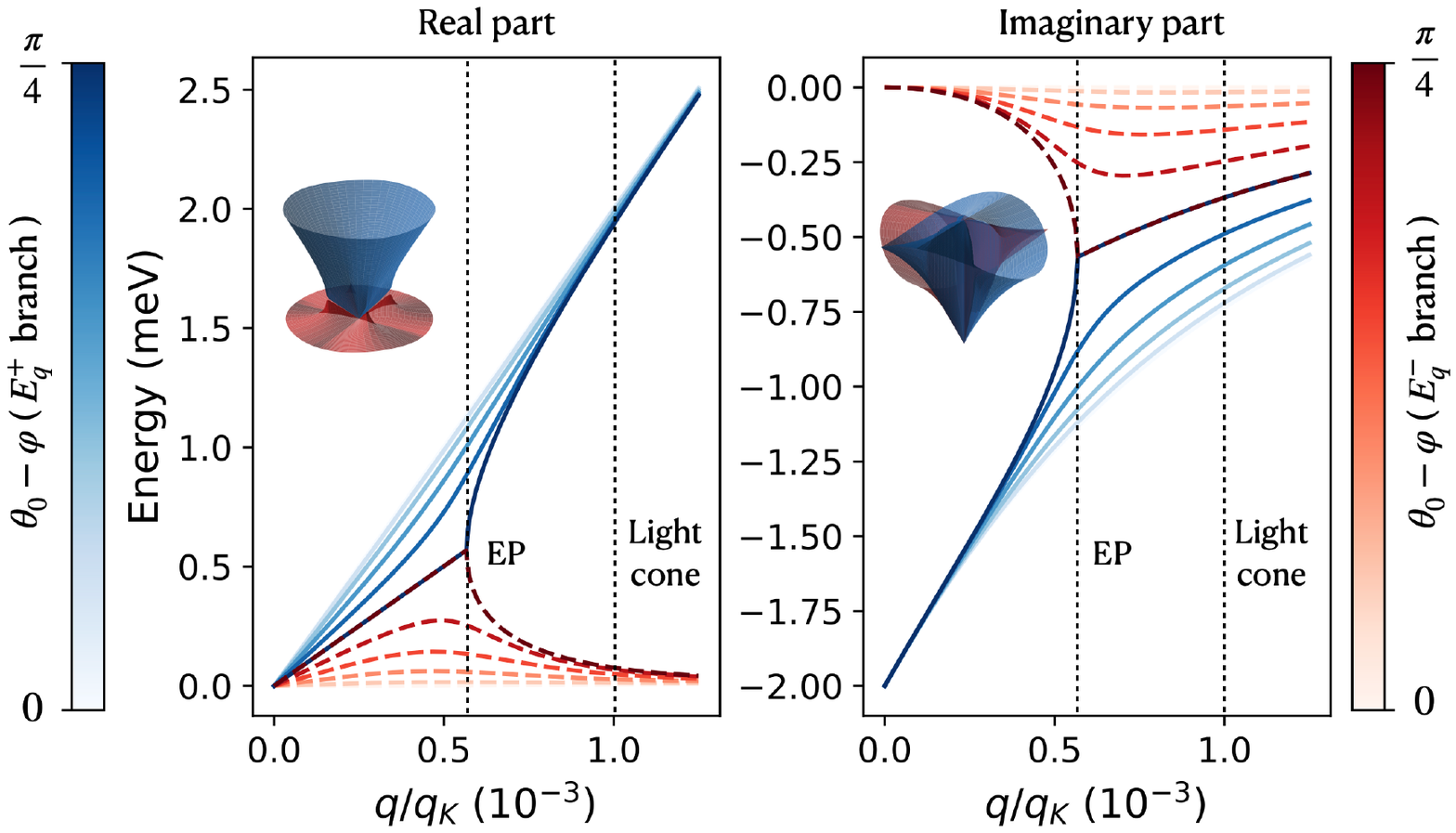}
\caption{\label{fig:ReE2} Excitonic spectrum for bath B0.
Real (left) and imaginary (right) 
part of the  eigenvalues 
$E^\pm_\blq$ as a function of the exciton momentum $q$ for 
different angles $\th_{0}-\vf$. The upper and lower 
branches are shown in solid-blue and dashed-red, respectively. At 
$\vf=\th_{0}\pm \pi/4$, 
an EP emerges. The insets 
show the three-dimensional dispersion, highlighting the anistropic 
structure in momentum space.}                 
\end{figure}

Figure~\ref{fig:ReE2} displays the excitonic 
eigenvalues in bath B0 for different values of the angle $\vf$.
The dissipation parameters are $D^{(0)}=1$~meV and $\l=1/q_{c}$.
Once again, the imaginary parts are both negative, as it should be.
Only excitons with momentum directions tilted by 
an angle of $\pm\frac{\pi}{4}$ with respect to the angle $\th_{0}$ of the 
linearly polarized photons become degenerate. The exceptional 
momentum satisfies $U\frac{q_{e}}{q_{K}}=D^{(0)}e^{-\l q_{e}}$, which 
yields $q_{e}\simeq 0.56 q_{c}$.  The 
excitonic spectrum exhibits two exceptional points rather than an 
exceptional ring.

\subsection{Valley polarization, light polarization and ellipticity}

\begin{figure}[tbp]
\includegraphics[width=0.48\textwidth]{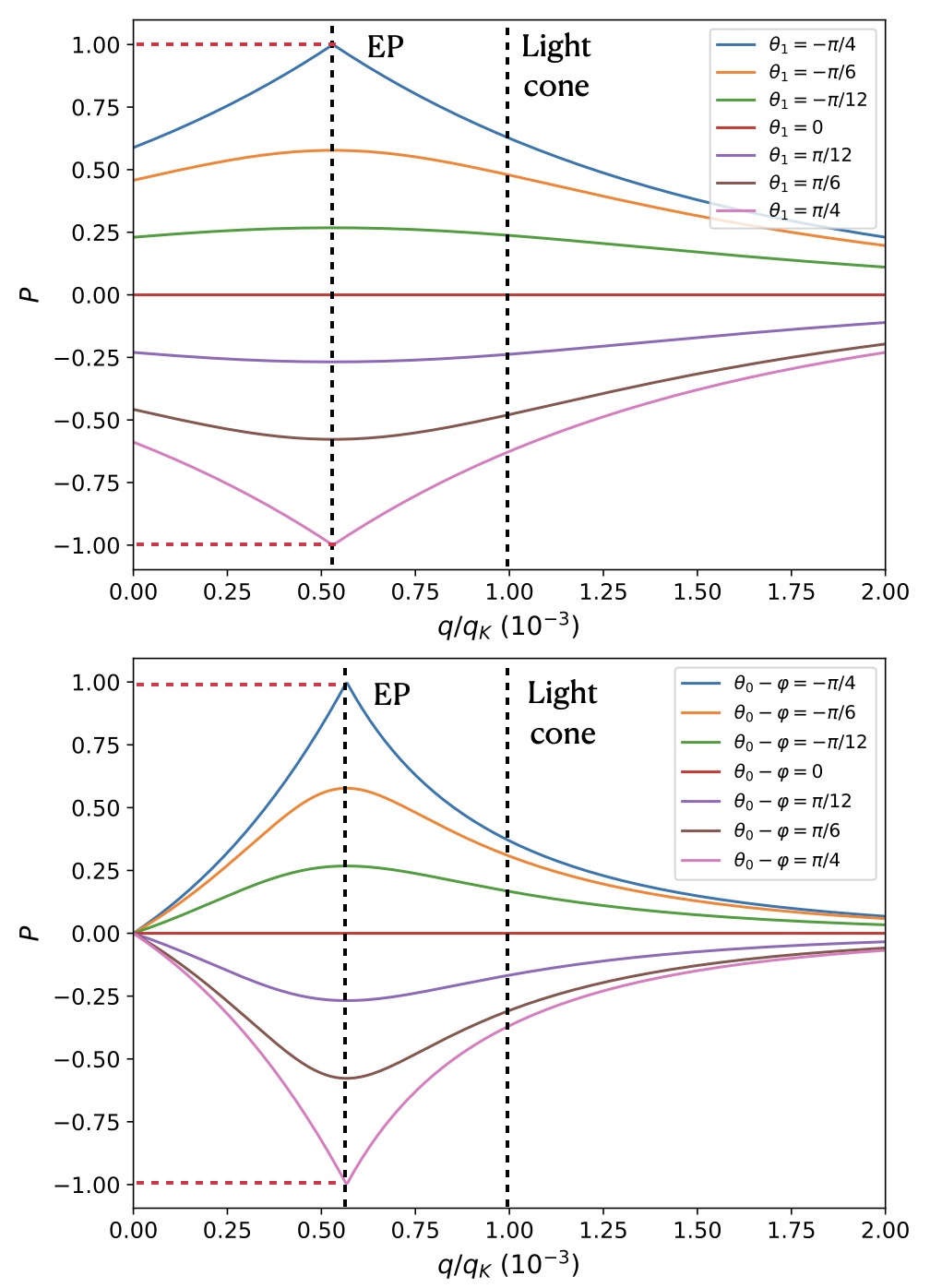}
\caption{\label{fig:valley_pol} Valley polarization 
$P_{\blq}$ as a function of momentum $q$ for bath B1 (top) and B0 
(bottom). For B1 different values of $\theta_{1}$ are considered. 
For $\theta_{1}=0$ the polarization 
vanishes at all momenta, reflecting perfect balance between the 
valleys. At 
finite $\theta_{1}$ the polarization reaches extremal 
values at the exceptional momentum. In particular, at 
$\theta_{1}=\pm\pi/4$ the valley polarization saturates, 
$P_{\blq}=\pm1$, indicating complete localization in either the $K$ or 
$K'$ valley. For B0 different values of the angle $\vf$ are considered. 
The valley polarization saturates at the exceptional momentum only 
for $\vf=\th_{0}\pm\p/4$. 
}                
\end{figure}

In this section, we study the valley polarization and characterize 
the polarization of light in optical emission.   

Let us write the normalized right  eigenvectors in Eq.~(\ref{eigenvector}) as 
$\vec{u}_{\blq,\rm R}^\pm=(A_{\blq},A'_{\blq})^{\top}$. 
The valley polarization 
\begin{align}
    P_\blq=|A_{\blq}|^2-|A'_{\blq}|^2,
\end{align}
measures the degree of circular  polarization in photoluminescence 
experiments~\cite{zeng2012valley,mak2012control,cao2012valley,jones2013optical}. 
Its value is independent of the branch, see Eqs.~(\ref{eigenvectorR}).
In Fig.~\ref{fig:valley_pol} we plot 
$P_\blq$ as a function of momentum $q$ for bath B1 (top) and B0 
(bottom). For B1  several representative values of the parameter 
$\theta_{1}$ are displayed. 
If $\th_{1}=0$  
the valley polarization is zero for all $\blq$. Any 
other choice of $\th_{1}$ breaks the balance between the $K$ and $K'$ valleys,  
with the polarization peaking at $\th_{1}=\pm \p/4$ near the exceptional 
momentum $q_e$ (vertical dashed line).
At the peak the polarization is unity, indicating that the 
system is fully polarized. This behavior directly links the appearance 
of EPs to the 
emergence of valley-selective excitonic states.   
A similar discussion can be made for bath B0, where the valley 
polarization is studied for different angles $\th_{0}-\vf$. In this 
case $P_{\blq=0}=0$ for all $\th_{0}$, as it should be. The polarization 
saturates at the exceptional momentum  $q_{e}$
only for momenta $\blq$ forming an angle $\vf=\th_{0}\pm\p/4$ with 
the $x$-axis.

Next we explore the momentum resolved light polarization in 
photoluminescence.
Let $\blJ=(J_x e^{{\rm i}\f_x},J_ye^{{\rm i}\f_y})^{\top}$
be the Jones vector describing a monochromatic electric field
propagating along the $z$-direction, which is perpendicular to the TMD plane.
The relative phase difference $\f_x-\f_y$ can be used to classify the 
light polarization. Specifically, the light is linearly polarized when
$|\f_x-\f_y|=0$, circularly polarized when 
$|\f_x-\f_y|=\pi /2$, and elliptically 
polarized for a general phase difference.      
Using the valley contrasting optical selection 
rules~\cite{yu2015valley,PhysRevB.77.235406,xiao2012coupled,schaibley2016valleytronics}, 
the Jones vector given by optical emission from the 
$\vec{u}_{\blq,\rm R}^\pm$ exciton has components   
$J_x e^{{\rm i}\f_x}=(A_{\blq}+A'_{\blq})/\sqrt{2}$ and 
$J_y e^{{\rm i}\f_y}={\rm i}(A_{\blq}-A'_{\blq})/\sqrt{2}$.

\begin{figure}[tbp]
\includegraphics[width=0.5\textwidth]{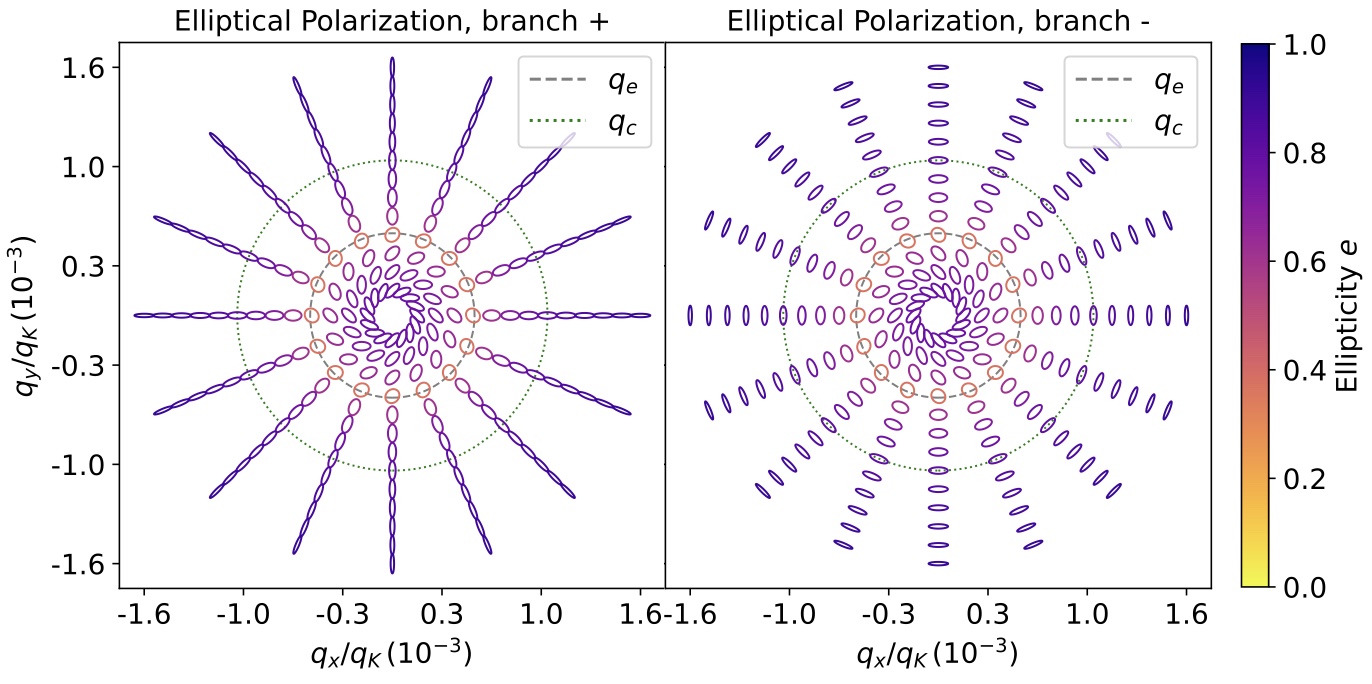}
\includegraphics[width=0.5\textwidth]{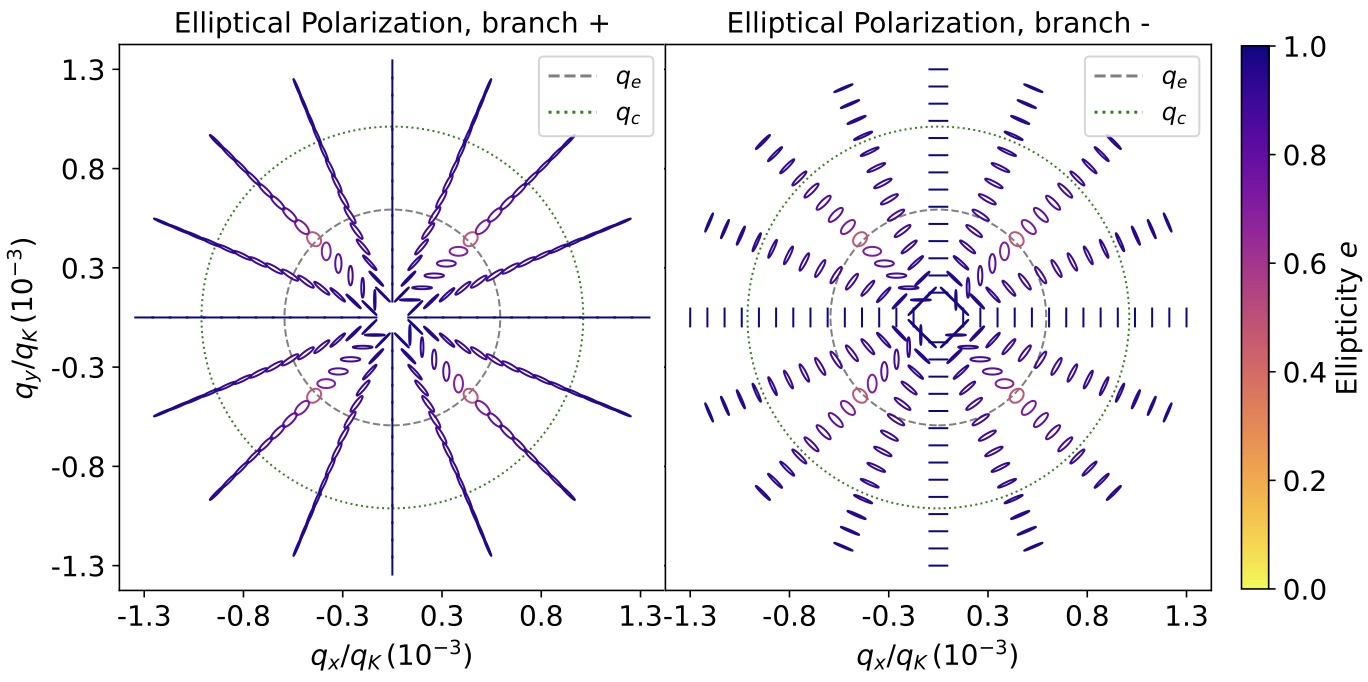}
\caption{\label{fig:ellipticity} Elliptical polarization patterns of 
the electric field $\blE(t)=\Re[\blJ e^{{\rm i}\w t}]$ 
for the $\pm$ branches and for bath B1 (top) and B0 (bottom). The 
orientation and shape of the ellipses encode the polarization state, 
while the color scale represents the ellipticity. At the 
EPs the electric field is circularly polarized, 
whereas for $q \to 0$ and $q \to \infty$ the polarization 
reduces to linear. The relative orientation of the polarization axes 
exhibits a $\pi/2$ difference between the two branches at large 
$q$.}                
\end{figure}

\begin{figure*}[tbp]
\includegraphics[width=0.99\textwidth]{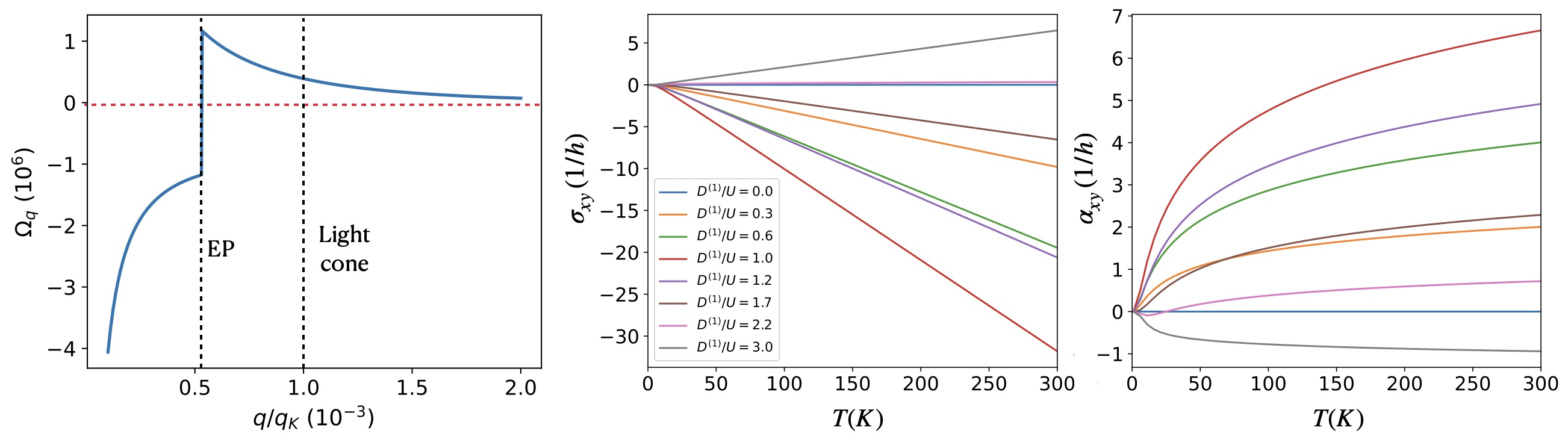}
\caption{(Left) Berry curvature $\Omega_{q}$ at 
$\theta_{1}=\pi/4$. The curve shows a divergence for $q \to 0$ 
and a discontinuous jump of magnitude $2\l^{2}\ln(D^{(1)}/U)$ 
across the exceptional momentum $q_e$, 
consistent with the analytic expressions in 
Eq.~(\ref{eq:berry_curv}). Middle-Right. Hall conductivity (middle), 
$\sigma_{xy}$, and Nernst conductivity (right), $\alpha_{xy}$, against 
temperature for several values of the ratio $D^{(1)}/U$. 
In both cases, the emergence of the exceptional ring at $D^{(1)}/U=1$ inverts the 
trend in how these quantities vary with  $D^{(1)}/U$. }             
\label{fig:Berry}
\end{figure*}

We first consider the bath B1, and fix 
$\th_{1}=\pi/4$. In Fig.~\ref{fig:ellipticity} (top) we
show how the electric field vector 
$\blE(t)=\Re[\blJ e^{{\rm i}\w t}]$ moves in the TMD plane over time for 
different $\blq$'s. For both $\pm$ branches, 
the elliptical polarization rotates as $\mathbf{q}$ varies, and the 
ellipticity changes continuously except at the EP, 
where circular polarization emerges. Notably, the orientation 
angle, defined as the angle between the major axis 
of the polarization ellipse and the $x$-axis, 
undergoes a sudden jump when crossing the EP. In the limit of 
large $q$, the two branches converge to orthogonal linear 
polarizations, consistent with the nondissipative (Hermitian) 
theory.

The polarization and ellipticity of the electric field 
$\blE(t)=\Re[\blJ e^{{\rm i}\w t}]$ for the bath B0 is diplayed in 
Fig.~\ref{fig:ellipticity} (bottom). Here, the photons of the bath are polarized 
along $x$, i.e., $\th_{0}=0$ (polarization along $y$ yields the same 
results). The anistropy is clearly visible. 
Circular polarization occurs only at the EPs, that is for 
$\vf=\pm \frac{\p}{4}$ 
and $\vf=\pm \frac{3\p}{4}$.

\subsection{Berry curvature and conductivities}

The non-Hermitian excitons in bath B1 have nontrivial topological 
properties.
The Berry connection from the normalized right eigenstates is
$\bcallC^\pm_\blq\equiv {\rm i}\vec{u}_{\blq,\rm R}^{\pm\dagger}
\grad_\blq\vec{u}_{\blq,\rm R}^\pm$~\cite{shen_topological_2018,wang2024non}, and its real part has vanishing 
radial component:
\begin{align}
    \Re[\bcallC^\pm_\blq]=\begin{cases}
        -\frac{1}{q}\frac{U_q}{D^{(1)}_q}\ble_\vf,\quad &q<q_e\\
        -\frac{1}{q}\frac{D^{(1)}_q}{U_q}\ble_\vf,\quad &q>q_e
    \end{cases}.
\end{align}
where $\ble_\vf$ is the angular unit vector (perpendicular to the 
radial direction). 
The associated Berry curvature for $\th_{1}=\p/4$ is independent of the 
branch and given by
\begin{align}
    \Omega_{q} \equiv \grad_\blq \times \Re [\bcallC^\pm_\blq]
=\begin{cases}
        -\frac{U}{D^{(1)}}\frac{\lambda e^{\lambda q}}{q},\quad &q<q_e\\
        +\frac{D^{(1)}}{U}\frac{\lambda e^{-\lambda q}}{q},\quad 
		&q>q_e
    \end{cases}.
\label{eq:berry_curv}	
\end{align}
Figure~\ref{fig:Berry} (left) 
highlights the singular nature of the Berry curvature in 
the vicinity of $q=0$, where $\Omega_{q}\sim -1/q$. The discontinuity 
of magnitude $2\l^{2}\ln(D^{(1)}/U)$
at the exceptional momentum $q=q_e$ marks a transition between two regimes with opposite 
analytical forms. For $q>q_e$, the curvature decays 
exponentially, ensuring that the Berry phase accumulated along a large 
loop remains finite.    

Using the rotational invariance of the eigenvalues in Eq.~(\ref{eigenvalue})
and the Berry curvature in Eq.~(\ref{eq:berry_curv}), the Hall 
conductivity and Nernst conductivity can be   
written as~\cite{wang2024non}
\begin{align}
    \s_{xy}&\equiv - \s_{0}
	\sum_{\t=\pm} 
	\int_0^\infty {\rm d} q f_\t^0(q)\left(q\W_q\right),\label{hall1}\\
	\a_{xy}&\equiv \frac{\s_{0}}{ T}\sum_{\t=\pm} 
	\int_0^\infty {\rm d} q \left(q\W_q\right)\nn\\
    &\times\Big[\frac{\Re[E_q^{\t}]-\m}{k_B}(f_\t^0(q)+1)+
	T\log f_\t^0(q)\Big]\label{nernst1},
\end{align}
where $\s_{0}=1/h$ is the quantum of conductance and
$f_\t^0(q)=1/[e^{ (\Re[E_q^\t]-\m)/k_BT}-1]$ is the 
bosonic exciton distribution. Both integrals are 
convergent and well-behaved.

Figure~\ref{fig:Berry} (middle and right) displayes the 
temperature dependence of the conductivities for chemical potential 
$\m=-1$~meV.   
The Hall conductivity depends almost linearly on $T$. Interestingly, 
for $D^{(1)}/U<1$ -- where there is no exceptional ring -- the ratio $\s_{xy}/T$ is 
negative and decreases as $D^{(1)}/U$ increases. The emergence of the exceptional ring for
$D^{(1)}/U>1$ inverts this trend: $\s_{xy}/T$ begins to increase with increasing
$D^{(1)}/U$, eventually turning positive. 
The exceptional ring has a similar impact on the 
Nernst conductivity. The quantity $\a_{xy}$ is positive and 
increases with  $D^{(1)}/U$ until the formation of the exceptional 
ring. Upon further increasing  $D^{(1)}/U$, the Nernst conductivity 
decreases and eventually becomes negative, 
reflecting the entropy-driven nature of the effect.
The parameter $D^{(1)}$ acts as a tuning knob, changing the sign of 
the conductivities.

\section{Conclusions}
\label{concsec}

In this work, we have developed a rigorous, microscopic framework for 
describing excitonic phenomena in open quantum systems. By extending the BSE to 
incorporate Lindbladian dissipation  within the NEGF formalism, 
we constructed a first-principles approach 
capable of capturing the interplay between many-body interactions and 
environmental coupling. This formalism goes beyond effective 
low-energy or phenomenological models by preserving the underlying 
fermionic structure of excitons and accounting for ab initio 
band structures and electron-hole interactions.  
As such, it can be integrated into 
existing ab initio BSE 
codes~\cite{MARINI20091392,DESLIPPE20121269,krause_implementation_2017,vorwerk_bethe-salpeter_2019,romero_abinit_2020},   enabling 
material-specific  predictions.  
An important outcome of our analysis is that not all forms of 
dissipation channels are permissible in the stationary 
limit. Stability and causality provide two key constraints for 
physically meaningful spectra. 

The formalism is applicable to a broad class of dissipative 
environments. In particular, we investigated baths 
composed of linearly and circularly polarized photons, and found that 
dissipation can give rise to exceptional points (EPs) in the 
excitonic band structure. Depending on the nature of the bath, these 
EPs may appear either as a finite set or as a ring-shaped continuum. 
The presence of EPs also leaves distinctive signatures 
in observable quantities, including nonanalytic behavior in the 
valley polarization, complex polarization patterns in 
photoluminescence, and a nontrivial topological character of the 
excitonic states. More generally, our results
establish a unifying first-principles framework that 
connects open quantum dynamics, excitonic physics, 
and non-Hermitian topology.

The methods developed here also provide a foundation for 
several exciting directions. The framework could be applied to study 
driven-dissipative phenomena, time-dependent protocols, or excitonic 
dynamics beyond the steady state. It also opens the possibility of 
including phonons from first-principles~\cite{stefanucci_in-and-out_2023}.
In fact, exciton-phonon interactions play a crucial role in 
converting coherent bright excitons into incoherent bright and dark 
excitons~\cite{selig_exciton_2016,thranhardt_quantum_2000,brem_exciton_2018,stefanucci_excitonic_2025}.        
Another active area of application for the non-Hermitian BSE is the 
study of exciton-polaritons in lossy cavities~\cite{su_direct_2021,deng_exciton-polariton_2010,davisson_simulating_2020,rosser_exciton-phonon_2020}.        
Finally, the appearance 
of topological features using engineered baths suggests 
opportunities for designing non-Hermitian quantum devices with 
controllable optical and valleytronic properties.

\acknowledgments

We acknowledge funding from Ministero Universit\`a e 
Ricerca PRIN under grant agreement No. 2022WZ8LME, 
from INFN through project TIME2QUEST, 
from European Research Council MSCA-ITN TIMES under grant agreement 101118915, 
and from Tor Vergata University through project TESLA.

\appendix

\section{Derivation of the dissipative Hamiltonian}
\label{lindHapp}

The starting point is the Hamiltonian in Eq.~(\ref{hzatez}). 
Inserting in $\hat{\mathcal{H}}(z)$ the explicit form of the jump operators in 
Eq.~(\ref{genlind}) we find (sum over repeated indices is implicit)   
\begin{widetext}
\begin{align}\label{H app expand}
    \hat{\mathcal{H}}(z)&=\hat H(z)-{\rm i}s(z)\Big[V_{ij}^{\rm loss}(t)\hat d_i^\dagger(z)\hat d_j(z)+V_{ij}^{\rm gain}(t)\hat d_j(z)\hat d_i^\dagger(z)+\frac{1}{2}v_{ijmn}^\mathrm{diss}(t)\hat{d}_i^\dagger(z)\hat d_n(z)\hat{d}_j^\dagger(z)\hat d_m(z)\Big]\nn\\
    &+2\mathrm{i}\theta_-(z)\Big[V_{ij}^{\rm loss}(t)\hat d_i^\dagger(z^\ast)\hat d_j(z)+\frac{1}{4}v_{ijmn}^{\rm diss}(t)\hat{d}_i^\dagger(z^\ast)\hat d_n(z^\ast)\hat{d}_j^\dagger(z)\hat d_m(z)\Big]\nn\\
    &-2\mathrm{i}\theta_+(z)\Big[V_{ij}^{\rm gain}(t)\hat 
	d_j(z)\hat d_i^\dagger(z^\ast)+\frac{1}{4}v_{jinm}^{\rm diss}(t)\hat{d}_j^\dagger(z)\hat d_m(z)\hat{d}_i^\dagger(z^*)\hat d_n(z^*)\Big],
\end{align}
where the quantities $V_{ij}^{\rm loss}$, $V_{ij}^{\rm gain}$ and 
$v_{jinm}^{\rm diss}$ are defined in Eqs.~(\ref{Vloss}-\ref{Vgain}) 
and Eq.~(\ref{dissint}) respectively.
Taking into account 
that under the $\callT$ sign the bosonic (fermionic) operators 
(anti)commute, the normal-ordered form of Eq.~(\ref{H app expand}) 
reads    
\begin{align}
    \hat{\mathcal{H}}(z)&=\hat H(z)-{\rm i}s(z)\Big[V_{ij}^{\rm loss}(t)\pm V_{ij}^{\rm gain}(t)+V_{ij}^{\rm diss}(t)\Big]\hat d_i^\dagger(z)\hat d_j(z)+2\mathrm{i}\Big[\theta_-(z)V_{ij}^{\rm loss}(t)\mp\theta_+(z)V_{ij}^{\rm gain}(t) \Big]\hat d_i^\dagger(z^\ast)\hat d_j(z)\nn\\
    &-\frac{\mathrm{i}}{2}s(z)v_{ijmn}^{\rm diss}(t)\hat{d}_i^\dagger(z)\hat{d}_j^\dagger(z)\hat d_m(z)\hat d_n(z)+\frac{\mathrm{i}}{2}
    \Big[\theta_-(z)v_{ijmn}^{\rm diss}(t)-\theta_+(z)v_{jinm}^{\rm diss}(t)\Big]\hat{d}_i^\dagger(z^\ast)\hat{d}_j^\dagger(z)\hat d_m(z)\hat d_n(z^\ast),
\end{align}
which can be verified to be the same as Eq.~(\ref{splitham}) once the 
integral over $\bar z'$ in Eqs.~(\ref{onebodyH}-\ref{twobodyH}) is 
performed.

\section{Martin-Schwinger Hierarchy for open systems}
\label{MSapp}
Let $\hat O_i(z_i) = f_i(z_i) \hat C_i(z_i)$, where $f_i(z_{i})$ is a 
scalar function depending on both $z_i$ and $z_i^\ast$, and $\hat 
C_i(z_i)$ is the product of an arbitrary number of field operators, 
some evaluated at $z_i$ and the remaining at $z_i^\ast$. 
The contour-time derivative of the contour-ordered string of such operators  is given by 
\begin{align}\label{eom in app}
    \frac{\rm d}{\mathrm{d}z_k}\mathcal{T}\big\{\hat O_1(z_1)\ldots\hat O_n(z_n)\big\}&=\sum_{l=1}^{k-1}(\pm)^{k-l}[\delta(z_k,z_l)+\delta(z_k,z_l^{\ast})]\mathcal{T}\big\{\hat O_1(z_1)\ldots\stackrel{\sqcap}{\hat O_l}\!(z_l)\ldots[\hat O_k(z_k),\hat O_l(z_l)]_\mp\ldots\hat O_n(z_n)\big\}\nn\\
    &+\sum_{l=k+1}^{n}(\pm)^{l-k-1}[\delta(z_k,z_l)+\delta(z_k,z_l^{\ast})]\mathcal{T}\big\{\hat O_1(z_1)\ldots[\hat O_k(z_k),\hat O_l(z_l)]_\mp\ldots \stackrel{\sqcap}{\hat O_l}\!(z_l)
    \ldots\hat O_n(z_n)\big\},
\end{align}
where the lower sign applies when both $\hat O_k$ and $\hat O_l$ are 
fermionic operators. We use this identity to derive the MSH. Taylor expanding the exponential of 
$\hat{\mathcal{H}}(z)$  in 
Eq.~(\ref{n-particleNEGF}), the $G_n$ reads   
\begin{align}\label{Gn for eom}
    G_n(j_1z_1,\ldots,j_nz_n;j_1'z_1',\ldots,j_n'z_n')=\frac{1}{\mathrm{i}^n}\sum_{k=0}^\infty\frac{(-\mathrm{i})^k}{k!}\int_C\mathrm{d}\bar{z}_1\ldots\mathrm{d}\bar z_k\Tr\Big[\hat \rho(t_0) \mathcal{T}\Big\{\hat{\mathcal{H}}(\bar z_1)\ldots\hat{\mathcal{H}}(\bar z_k)\hat d_{j_1}(z_1)\ldots\hat d_{j_1'}^\dagger(z_1')\Big\}\Big].
\end{align}
Taking the contour-time derivative of Eq.~(\ref{Gn for eom}) with respect to $z_q$ 
(corresponding to field operator $\hat d_{j_q}(z_q)$), 
using Eq.~(\ref{eom in app}), and performing the integrals over the 
delta functions arising from the commutators with $\hat{\mathcal{H}}$, 
the equation of motion for $G_n$ follows
\begin{align}\label{msh app1}
    &\mathrm{i}\frac{\rm d}{\mathrm{d}z_q}G_n(j_1z_1,\ldots ,{j_q}{z_q},\ldots ,j_nz_n;j_1'z_1',\ldots,j_n'z_n')\nn\\
    &=\sum_{l=1}^n(\pm)^{q+l}\delta(z_q,z_l')\delta_{j_qj_l'}\,G_{n-1}(j_1z_1,\ldots ,\stackrel{\sqcap}{j_qz_q},\ldots ,j_nz_n;j_1'z_1',\ldots ,\stackrel{\sqcap}{j_l'z_l'},\ldots ,j_n'z_n')\nn\\
    &+\frac{1}{\mathrm{i}^n}\Big\langle \hat 
	d_{j_1}(z_1)\ldots\Big\{\big[\hat d_{j_q}(z_q),\hat{\mathcal{H}}(z_q)\big]_--\big[\hat d_{j_q}(z_q),\hat{\mathcal{H}}(z_q^*)\big]_-\Big\}\ldots\hat d_{j_1'}^\dagger\Big\rangle.
\end{align}
\end{widetext}
To prove the equivalence of this result with the MSH we need 
to evaluate the
commutator between $\hat d_{j_q}(z_q)$ and Hamiltonians
$\hat{\mathcal{H}}(z_q)$ and 
$\hat{\mathcal{H}}(z_q^*)$. 
Let us introduce the quantities 
\begin{subequations}
\begin{align}
    U_{ij}(z)&=h_{ij}(t)
    -\mathrm{i}s(z)\Big[V_{ij}^{\rm loss}(t)\pm V_{ij}^{\rm gain}(t)+V_{ij}^{\rm diss}(t)\Big],
	\label{U from h}\\
    W_{ij}(z)&=2\mathrm{i}\Big[\theta_-(z)V_{ij}^{\rm loss}(t)\mp\theta_+(z)V_{ij}^{\rm gain}(t) \Big]
	\label{W from h},\\
    u_{ijmn}(z)&=v_{ijmn}^{\rm phys}(t)-\mathrm{i}s(z)v_{ijmn}^{\rm s,diss}(t)
	\label{u from v},\\
    w_{ijmn}(z)&=-\mathrm{i}s(z)v_{ijmn}^{\rm 
	s,diss}(t)-\mathrm{i}v_{ijmn}^{\rm a,diss}(t),
    \label{w from v}
\end{align}
\label{UWuw app}
\end{subequations}
and rewrite Eqs.~(\ref{oneph}) and (\ref{vdiss}) as
\begin{align}\label{h by U and W}
    h_{ij}(z',z)=\delta(z',z)U_{ij}(z)+\delta(z',z^*)W_{ij}(z),
\end{align} 
and 
\begin{align}\label{v by u and w}
    v_{ijmn}(z,z')=\delta(z,z')u_{ijmn}(z)+\delta(z,z^{\prime 
	*})w_{ijmn}(z).
\end{align}
Inserting these results into Eqs.~(\ref{onebodyH}-\ref{twobodyH}) and 
performing the integral over $\bar z'$ we obtain (sum over repeated 
indices is implicit)
\begin{align}
    \hat{\mathcal{H}}(z)&=U_{ij}(z)\hat d_i^\dagger(z)\hat d _j(z)+W_{ij}(z)\hat d_i^\dagger(z^*)\hat d _j(z)\nn\\
    &+\frac{1}{2}u_{ijmn}(z)\hat{d}_i^\dagger(z)\hat{d}_j^\dagger(z)\hat d_m(z)\hat d_n(z)\nn\\
    &-\frac{1}{2}w_{ijmn}(z)\hat{d}_i^\dagger(z^\ast)\hat{d}_j^\dagger(z)\hat d_m(z)\hat d_n(z^\ast).
\end{align}
Under the contour-ordering sign we have the following 
identities [field operators on 
different branches (anti)commute]     
\begin{align}\label{com dq Hz}
    \big[\hat d_q(z),\hat{\mathcal{H}}(z)\big]_-&=U_{qn}(z)\hat d_n(z)+u_{jqnm}(z)\hat d_j^\dagger(z)\hat d_m(z)\hat d_n(z)\nn\\
    &-\frac{1}{2}w_{jqnm}(z)\hat d_j^\dagger(z^*)\hat d_m(z^*)\hat 
	d_n(z),
\end{align}
and
\begin{align}\label{com dq Hz*}
    \big[\hat d_q(z),\hat{\mathcal{H}}(z^*)\big]_-&=W_{qn}(z^*)\hat d_n(z^*)\nn\\
    &-\frac{1}{2}w_{qjmn}(z^*)\hat d_j^\dagger(z^*)\hat d_m(z^*)\hat d_n(z).
\end{align}
Subtracting Eq.~(\ref{com dq Hz*}) from Eq.~(\ref{com dq Hz}) 
yields
\begin{align}\label{dH-dH*}
    &\big[\hat d_{j_q}(z_q),\hat{\mathcal{H}}(z_q)\big]_--\big[\hat d_{j_q}(z_q),\hat{\mathcal{H}}(z_q^*)\big]_-\nn\\
    &=\int_C\mathrm{d}z'h_{j_qn}(z,z')\hat d _n(z')\nn\\
    &+\int_C \mathrm{d}z'v_{jj_qnm}(z,z')\hat d_j^\dagger(z')\hat 
	d_m(z')\hat d_n(z).
\end{align}
Substituting Eq.~(\ref{dH-dH*}) into Eq.~(\ref{msh app1}) we 
recover the first set of equations of the MSH, see Eq.~(\ref{mshL}).
The second set of equations can be 
derived in a similar way.      

\section{Hartree-Fock one-particle Hamiltonian and Green's function}
\label{HFapp}
We first prove that $\int_C 
\mathrm{d}z\mathrm{d}z'[h(z,z')+\Sigma^{\rm HF}(z,z')]=0$. From 
Eqs.~(\ref{oneph}) and (\ref{hartree}), it is easy to verify 
that    
\begin{align}
    \int_C\mathrm{d}z\mathrm{d}z'h_{in}(z,z')&=-2\mathrm{i}\int_{t_0}^\infty\mathrm{d}\bar t \,V_{in}^{\rm diss}(\bar t),\\
    \int_C\mathrm{d}z\mathrm{d}z'\Sigma_{in}^{\rm H}(z,z')&=0.
\end{align}
We evaluate the contour integral of the Fock self-energy using 
Eqs.~(\ref{generalized v}-\ref{vdiss}) and Eq.~(\ref{fock}). We find  
(sum over repeated 
indices is implicit)
\begin{align}
    \int_C\mathrm{d}z\mathrm{d}z'\Sigma_{in}^{\rm F}(z,z')&=\int_C\mathrm{d}z\Big\{s(z)G^<_{mj}(t,t)v_{ijnm}^{\rm s,diss}(t)\nn\\
    &+G_{mj}(z,z^*)\Big[s(z^*)v_{ijnm}^{\rm s,diss}(t)+v_{ijnm}^{\rm a,diss}(t)\Big]\Big\}
	\nn\\
		&=\int_{t_0}^\infty \mathrm{d}t\big[G^<_{mj}(t,t)-G^>_{mj}(t,t)\big]v_{ijnm}^{\rm diss}(t)\nn\\
    &=2\mathrm{i}\int_{t_0}^\infty\mathrm{d}t V_{in}^{\rm diss}(t),
\end{align}
where we took into account that $G(t_\mp,t'_{\pm})=G^\lessgtr(t,t')$. 
We conclude that stability is recovered  through the Fock self-energy.

In order to derive the solution of the HF KBE, we split 
the self-energy as $\Sigma^{\rm s}=\Sigma^{\rm HF}+\Sigma^{\rm c}$. 
The HF approximation corresponds to set $\Sigma^{\rm c}=0$. Thus 
the equation of motion in Eq.~(\ref{eom1}) becomes
\begin{align}\label{eom1 in apdx}
    {\rm i}\frac{{\rm d}}{{\rm d}z}G(z,z')&-\int_C {\rm d}\bar{z} 
	\left[h(z,\bar{z})+\Sigma^{\rm 
	HF}(z,\bar{z})\right]G(\bar{z},z')=\delta(z,z').
\end{align}
We rewrite the HF 
Hamiltonian in the form
\begin{align}\label{h+hf=U+W}
    h(z,\bar{z})+\Sigma^{\rm HF}(z,\bar{z})=\delta(z,\bar{z})U^{\rm 
	HF}(\bar{z})+\delta(z,\bar{z}^*)W^{\rm HF}(\bar{z}). 
\end{align}
In terms of the quantities introduced in 
Eqs.~(\ref{h by U and W}-\ref{v by u and w}) we have 
\begin{align}
    U_{in}^{\rm HF}(\bar{z})&=U_{in}(\bar z)+V_{in}^{\rm HF}(\bar 
	t)+V_{in}^{\rm a,diss}(\bar t)
	\nn\\&\mp\mathrm{i}s(\bar z)v_{ijnm}^{\rm s,diss}(\bar t)\rho_{mj}^<(\bar t),\\
    W_{in}^{\rm HF}(\bar z)&=W_{in}(\bar z)-\mathrm{i}G_{mj}(\bar z^*,\bar z)w_{ijnm}(\bar z).
\end{align}
Substituting Eq.~(\ref{h+hf=U+W}) into Eq.~(\ref{eom1 in apdx}) and 
taking the integral over $\bar z$ we obtain
\begin{align}
    &\left[{\rm i}\frac{{\rm d}}{{\rm d}z}-U^{\rm HF}(z)\right]G(z,z')+
	W^{\rm HF}(z^*)G(z^*,z')=\delta(z,z').
\end{align}
The HF KBE for the lesser component follows when setting $z=t_-$ and 
$z'=t_+'$, and is identical to Eq.~(\ref{kbe1}) with $\Sigma^{\rm c}=0$. 
The other HF KBEs can be derived in a similar way.   

Let us prove that the HF KBEs admit the following solution 
\begin{align}
    G^{\rm HF}(z,z')&=\th(t-t')T\Big\{e^{-{\rm i}\int_{t'}^t {\rm 
	d}\bar{t}h^{\rm HF}(\bar{t})}\Big\} G^{\rm HF}(z^{\prime *},z')\nn\\
    &+\theta(t'-t)G^{\rm HF}(z,z^*)\bar{T}\Big\{e^{+{\rm i}\int_{t}^{t'} {\rm d}\bar{t}h^{\rm HF\dagger}(\bar{t})}\Big\},
\end{align}
according to which the lesser and greater components read
\begin{align}\label{G HF <> apdx}
    G^{\rm HF,\lessgtr}(t,t')&=\theta(t-t')T\left\{e^{-\mathrm{i}\int_{t'}^t \mathrm{d}\bar{t}h^{\rm HF}(\bar{t})}\right\} G^{\rm HF,\lessgtr}(t',t')\nn\\
    &+\theta(t'-t)G^{\rm 
	HF,\lessgtr}(t,t)\bar{T}\left\{e^{+\mathrm{i}\int_{t}^{t'} \mathrm{d}\bar{t}h^{\rm HF\dagger}(\bar{t})}\right\},
\end{align}
and, consequently,  the retarded and advanced components are given by
\begin{align}
    G^{\rm HF,R}(t,t')&=-\mathrm{i}\th(t-t') T\left\{e^{-\mathrm{i}\int_{t'}^t 
	\mathrm{d}\bar{t}h^{\rm HF}(\bar{t})}\right\},\label{G HF R apdx}\\
    G^{\rm HF,A}(t,t')&=+\mathrm{i}\th(t'-t) 
	T\left\{e^{+\mathrm{i}\int_{t}^{t'} d\bar{t}h^{\rm 
	HF\dagger}(\bar{t})}\right\}.\label{G HF A apdx}
\end{align}

Taking the derivative of $G^<(t,t')$ in Eq.~(\ref{G HF <> apdx})
with respect to the real time $t$ we find
\begin{align}\label{eom G< hf apdx}
    &\mathrm{i}\frac{\rm d}{\mathrm{d}t}G^{\rm HF,<}(t,t')\nn\\
    &=\theta(t-t')h^{\rm HF}(t)T\left\{e^{-\mathrm{i}\int_{t'}^t \mathrm{d}\bar{t}h^{\rm HF}(\bar{t})}\right\} G^{\rm HF,<}(t',t')\nn\\
    &+\theta(t'-t)\left[\mathrm{i}\frac{\rm d}{\mathrm{d}t}G^{\rm HF,<}(t,t)\right]\bar{T}\left\{e^{+\mathrm{i}\int_{t}^{t'} \mathrm{d}\bar{t}h^{\rm HF\dagger}(\bar{t})}\right\}\nn\\
    &+\theta(t'-t)G^{\rm HF,<}(t,t)h^{\rm HF\dagger}(t)\bar{T}\left\{e^{+\mathrm{i}\int_{t}^{t'} \mathrm{d}\bar{t}h^{\rm HF\dagger}(\bar{t})}\right\}.
\end{align}
The equation of motion for equal time $G^{\rm HF,<}(t,t)$ -- appearing 
in third line of the above equation -- can be expressed through Eq.~(\ref{lyapunov}) (with 
collision integral $I(t)=0$): 
\begin{align}\label{G<(tt) HF}
    \mathrm{i}\frac{\rm d}{\mathrm{d}t}G^{\rm HF,<}(t,t)&=
	h^{\rm HF}(t)G^{\rm HF,<}(t,t)-G^{\rm HF,<}(t,t)h^{\rm HF\dagger}(t)
	\nn\\&\pm 2\gamma^<(t).
\end{align}
Substituting Eq.~(\ref{G<(tt) HF}) into the third line of 
Eq.~(\ref{eom G< hf apdx}) and identifying the expressions of $G^{\rm 
HF,<}$ and $G^{\rm HF,A}$ in Eqs.~(\ref{G HF <> apdx}) and (\ref{G HF A 
apdx}), we find   
\begin{align}
    \mathrm{i}\frac{\rm d}{\mathrm{d}t}G^{\rm HF,<}(t,t')=h^{\rm 
	HF}(t)G^{\rm HF,<}(t,t')\mp 2i\gamma^<(t)G^{\rm HF,A}(t,t'),
\end{align}
which correctly agrees with the HF KBE in Eq.~(\ref{kbe1}). The other KBEs can be verified 
in a similar way.

Finally, we consider the stationary case of a one-particle density matrix 
that satisfies the Lyapunov equation with 
time-independent $\hat H$ and $\hat{\mathcal{D}}$, see Eq.~(\ref{lyapunov}), 
\begin{align}
-{\rm i}h^{\rm HF}\rho^{<}+
{\rm i}\rho^{<}h^{\rm HF\dagger}+2{\g}^<=0.
\end{align}
We specialize the discussion to Hermitian matrices 
$(h^{\rm HF}+h^{\rm HF\dagger})$, $\g^{\gtrless}$ 
and $\rho^{<}$ that can be simulataneously 
diagonalized. We replace HF with qp, as any quasi-particle 
approximation with the same properties leads to identical results.  
In the basis that diagonalizes these matrices we have 
$h^{\rm qp}_{in}=\delta_{in}\left(\epsilon_i-{\rm i}\gamma_i\right)
$, and the stationary $\rho^{<}_{in}=\delta_{in}f_i$ that solves 
the Lyapunov equation is given by 
$f_i=\g^{<}_{i}/\g_{i}$.    
From Eqs.~(\ref{G HF <> apdx}) and (\ref{G HF R apdx}-\ref{G HF A apdx}) 
we then have the following Keldysh components for the quasiparticle NEGF
\begin{subequations} \label{qp g} 
\begin{align}
    G^{\rm qp,R}_{mn}(t_1,t_2)=&-{\rm i}\d_{mn}\th(t_1-t_2) e^{-{\rm i}(\e_m-{\rm i}\g_m)(t_1-t_2)},\\
    G^{\rm qp,A}_{mn}(t_1,t_2)=&+{\rm i}\d_{mn}\th(t_2-t_1) e^{+{\rm i}(\e_m+{\rm i}\g_m)(t_2-t_1))},\\
    G^{\rm qp,<}_{mn}(t_1,t_2)=&\mp {\rm i}\d_{mn} f_m\Big[\th(t_1-t_2)e^{-{\rm i}(\e_m-{\rm i}\g_m)(t_1-t_2)}\nn\\
    &+\th(t_2-t_1)e^{+{\rm i}(\e_m+{\rm i}\g_m)(t_2-t_1)}\Big],\\
    G^{\rm qp,>}_{mn}(t_1,t_2)=&- {\rm i}\d_{mn} \bar{f}_m\Big[\th(t_1-t_2)e^{-{\rm i}(\e_m-{\rm i}\g_m)(t_1-t_2)}\nn\\
    &+\th(t_2-t_1)e^{+{\rm i}(\e_m+{\rm i}\g_m)(t_2-t_1)}\Big],
\end{align}
\end{subequations}
where $\bar{f}\equiv 1\pm f$. Extracting the retarded component in 
Eq.~(\ref{def ell}) and using Eqs.~(\ref{qp g}) we also have
\begin{align}\label{ell r}
    \ell_{\substack{ij\\qp}}^{\rm qp,R}(t_1,t_2)=&\d_{ip}\d_{qj}\th(t_1-t_2)(f_j-f_i)\nn\\
    &e^{-{\rm i}[\e_i-\e_j-{\rm i}(\g_i+\g_j)](t_1-t_2)}.
\end{align}
The dissipation gives rise to a finite quasi-particle 
lifetime even in the absence of interactions.

\section{Retarded Bethe-Salpeter equation}
\label{retBSEapp}
From Eq.~(\ref{defxcl}), the anomalous and normal XC functions are 
given by
\begin{align}\label{appen L*}
    \pm L(d'z_b'^*,bz_b;b'z_b',dz_b)&=-\big\langle\hat d_{b'}^\dagger(z_b'^+)\hat d_{d'}(z_b'^*)\hat d_d^\dagger(z_b^+)\hat d_b(z_b)\big\rangle\nn\\
    &-G_{d'b'}(z_b'^*,z_b')G^<_{bd}(t_b,t_b),
\end{align}
and
\begin{align}\label{appen L}
    \pm L(d'z_b',bz_b;b'z_b',dz_b)&=-\big\langle\hat d_{b'}^\dagger(z_b'^+)\hat d_{d'}(z_b')\hat d_d^\dagger(z_b^+)\hat d_b(z_b)\big\rangle\nn\\
    &-G^<_{d'b'}(t_b',t_b')G^<_{bd}(t_b,t_b).
\end{align}
We consider the physical time $t_b'>t_b$, set $z_b'=t_{b-}'$ and 
$z_b=t_{b-}$, and subtract Eq.~(\ref{appen L}) from (\ref{appen L*}). 
We find the identity
\begin{align}
    &\pm \big[L(d't_{b+}',bt_{b-};b't_{b-}',dt_{b-})-L(d't_{b-}',bt_{b-};b't_{b-}',dt_{b-})\big]\nn\\
    &=-\big[\hat{d}_{b'}^\dagger,\hat{d}_{d'}\big]_\mp\langle\hat d_d^\dagger(t_{b-})\hat d_b(t_{b-})\rangle\nn\\
    &-[G^<_{d'b'}(t_b',t_b')-G^>_{d'b'}(t_b',t_b')]G^<_{bd}(t_b,t_b)=0,
\end{align}
where we use that  
$\langle\hat d_d^\dagger(t_{b-})\hat d_b(t_{b-})\rangle=\pm 
\mathrm{i}G^<_{bd}(t_b,t_b)$ along with  
$G^<(t,t)-G^>(t,t)=\mathrm{i}$.  Similarly, we have    
for $z_b'=t_{b-}'$ and $z_b=t_{b+}$
\begin{align}
    &\pm \big[L(d't_{b+}',bt_{b+};b't_{b-}',dt_{b+})-L(d't_{b-}',bt_{b+};b't_{b-}',dt_{b+}))\big]\nn\\
    &=-\big[\hat{d}_{b'}^\dagger,\hat{d}_{d'}\big]_\mp\langle\hat d_d^\dagger(t_{b+})\hat d_b(t_{b+})\rangle\nn\\
    &-[G^<_{d'b'}(t_b',t_b')-G^>_{d'b'}(t_b',t_b')]G^<_{bd}(t_b,t_b)=0.
\end{align}
In the same way we can show that the difference between the normal 
and anomalous XC functions is zero 
when $z_b'=t_{b+}'$ and $z_b=t_{b+}$ as well as when
$z_b'=t_{b+}'$ and $z_b=t_{b-}$. Thus, the equivalence in Eq.~(\ref{L*=L}) 
for all the physical times $t_b'>t_b$ is established.

Setting $z_c=z_a=z_{1}$ and $z_d=z_b=z_{2}$ in Eq.~(\ref{bse static 
hsex}), and renaming the indices $a\to i$, $b\to m$, $c\to j$ and 
$d\to n$, we obtain 
\begin{align}
    L_{\substack{ij\\nm}}(z_1,z_2)&=\ell_{\substack{ij\\nm}}(z_1,z_2)+\mathcal{A}_{\substack{ij\\nm}}(z_1,z_2)\nn\\
    &+\mathcal{B}_{\substack{ij\\nm}}(z_1,z_2)+\mathcal{C}_{\substack{ij\\nm}}(z_1,z_2),
\end{align}
where (sum over repeated indices is implicit)
\begin{widetext}
\begin{subequations}
\begin{align}
    \mathcal{A}_{\substack{ij\\nm}}(z_1,z_2)&=\mathrm{i}\int_C\mathrm{d}\bar z\,\ell_{\substack{ij\\pq}}(z_1,\bar z)\Big[ W_{psqr}^{\rm phys}(\bar{t})\pm v_{psrq}^{\rm phys}(\bar t)-\mathrm{i}s(\bar z)\left(v_{psqr}^{\rm s,diss}(\bar{t})\pm v_{psrq}^{\rm s,diss}(\bar t)\right)\Big]L_{\substack{rs\\nm}}(\bar z, z_2)
    \label{apdx A},\\
    \mathcal{B}_{\substack{ij\\nm}}(z_1,z_2)&=\mathrm{i}\int_C\mathrm{d}\bar z\,
    G_{ip}(z_1,\bar z^*)G_{qj}(\bar z,z_1)\left[\mathrm{i}s(\bar z)v_{psqr}^{\rm s,diss}(\bar t)+\mathrm{i}v_{psqr}^{\rm a,diss}(\bar t)\right]L(r\bar z^*,mz_2;s\bar z,nz_2^+)
    \label{apdx B},\\
    \mathcal{C}_{\substack{ij\\nm}}(z_1,z_2)&=\pm\mathrm{i}\int_C\mathrm{d}\bar z\,
    G_{ip}(z_1,\bar z^*)G_{qj}(\bar z^*,z_1)\left[\mathrm{i}s(\bar 
	z)v_{psrq}^{\rm s,diss}(\bar t)+\mathrm{i}v_{psrq}^{\rm 
	a,diss}(\bar t)\right]L_{\substack{rs\\nm}}(\bar z, z_2)\label{apdx C}.
\end{align}
\label{apdx ABC}
\end{subequations}
By definition, to extract the retarded component, 
we must evaluate $\callA$, $\callB$ and $\callC$ for a physical time $t_1>t_2$. 
We can replace the domain of integration $C$ by the shrunken contour 
$\tilde C=(t_2,t_1)\cup (t_1,t_2)$ since all integrals 
over $C/\tilde{C}$ are zeros. This allows as 
to replace $L(r\bar z^*,mz_2;s\bar z,nz_2^+)$ in Eq.~(\ref{apdx B}) 
by $L_{\substack{rs\\mn}}(\bar z, z_2)$. Consequently, we find
\begin{subequations}
\begin{align}
\label{callA apdx}
\mathcal{A}_{\substack{ij\\nm}}^{\rm 
R}(t_1,t_2)&=\mathrm{i}\int_{t_2}^{t_1}\mathrm{d}\bar 
t\Big\{\ell_{\substack{ij\\pq}}^{\rm R}(t_1,\bar 
t)\Big[W_{psqr}^{\rm phys}(\bar{t})\pm v_{psrq}^{\rm phys}(\bar 
t)\Big]-\mathrm{i}\Big[\ell_{\substack{ij\\pq}}^{>}(t_1,\bar 
t)+\ell_{\substack{ij\\pq}}^{<}(t_1,\bar t)\Big]\Big[v_{psqr}^{\rm 
s,diss}(\bar{t})\pm v_{psrq}^{\rm s,diss}(\bar 
t)\Big]\Big\}L_{\substack{rs\\nm}}^{\rm R}(\bar t,t_2)       
\\
\label{callB apdx}
\mathcal{B}_{\substack{ij\\nm}}^{\rm 
R}(t_1,t_2)&=\mathrm{i}\int_{t_2}^{t_1}\mathrm{d}\bar 
t\Big\{\mathrm{i}G_{ip}^<(t_1,\bar t)G_{qj}^<(\bar 
t,t_1)v_{psqr}^{\rm diss}(\bar{t})+\mathrm{i}G_{ip}^>(t_1,\bar 
t)G_{qj}^>(\bar t,t_1)v_{sprq}^{\rm 
diss}(\bar{t})\Big\}L_{\substack{rs\\nm}}^{\rm R}(\bar t,t_2),       
\\
\label{callC apdx}
\mathcal{C}_{\substack{ij\\nm}}^{\rm 
R}(t_1,t_2)&=\mp\mathrm{i}\int_{t_2}^{t_1}\mathrm{d}\bar 
t\,\Big\{\mathrm{i}\ell_{\substack{ij\\pq}}^{\rm R}(t_1,\bar 
t)v_{psrq}^{\rm a,diss}(\bar 
t)-\mathrm{i}\Big[\ell_{\substack{ij\\pq}}^{>}(t_1,\bar 
t)+\ell_{\substack{ij\\pq}}^{<}(t_1,\bar t)\Big]v_{psrq}^{\rm 
s,diss}(\bar t)      
\Big\}L_{\substack{rs\\nm}}^{\rm R}(\bar t,t_2).
\end{align}
\label{abcret}
\end{subequations}
Substituting Eqs.~(\ref{abcret}) into 
$L_{\substack{ij\\nm}}^{\rm R}(t_1,t_2)=\ell_{\substack{ij\\nm}}^{\rm 
R}(t_1,t_2)+\mathcal{A}_{\substack{ij\\nm}}^{\rm 
R}(t_1,t_2)+\mathcal{B}_{\substack{ij\\nm}}^{\rm 
R}(t_1,t_2)+\mathcal{C}_{\substack{ij\\nm}}^{\rm R}(t_1,t_2)$, 
and using $\ell_{\substack{ij\\pq}}^{\lessgtr}(t_1,\bar 
t)=G_{ip}^{\lessgtr}(t_1,\bar t)G^\gtrless_{qj}(\bar t,t)$ from 
Eq.~(\ref{def ell}) we recover Eq.~(\ref{bse lr}).     
\end{widetext}


%

\end{document}